\documentclass{jfm}

\usepackage{verbatim}
\usepackage{lscape}
\usepackage{booktabs}
\usepackage{amsmath,bm}
\usepackage{placeins}
\usepackage[utf8]{inputenc}
\usepackage{amsfonts}
\usepackage{amssymb}
\usepackage{xcolor}
\usepackage{filecontents}
\usepackage{physics}
\usepackage{float}
\usepackage{mathtools}
\usepackage{lipsum}
\usepackage{enumitem}
\usepackage{pgf}
\usepackage{pgfplots}
\pgfplotsset{width=2.4in, height=1.9in,every axis plot/.append style={thick}, compat=1.9} 
\newcommand\uu{{\bf u}}
\renewcommand\qq{{\bf q}}

\newcommand{\sprime}{^{\prime}}
\newcommand{\dprime}{^{\prime\prime}}
\newcommand\pd{{\partial }}

\newcommand{\nab}{{\bm\nabla}}

\newcommand{\Ra}{\RA}
\newcommand{\Nu}{\NU}

\newcommand\NU{\mbox{\textit{Nu}}}
\newcommand\RA{\mbox{\textit{Ra}}}

\renewcommand{\vec}[1]{\mathbf{#1}}

\definecolor{gfblue4}{RGB}{0, 102, 155}
\definecolor{gfblue3}{RGB}{0, 137, 204}
\definecolor{gfblue2}{RGB}{7, 170, 255}
\definecolor{gfblue1}{RGB}{130, 211, 255}
\definecolor{gfred1}{RGB}{255, 117, 186}
\definecolor{gfred2}{RGB}{255, 0, 124}
\definecolor{gfred3}{RGB}{196, 0, 96}
\definecolor{gfred4}{RGB}{153, 0, 76}

\definecolor{gfblue}{RGB}{0, 102, 156}
\definecolor{gfred}{RGB}{154, 0, 77}

\title{Flow states and heat transport in Rayleigh--B\'enard convection with different sidewall boundary conditions}

\author{Philipp Reiter\aff{1}\corresp{\email{philipp.reiter@ds.mpg.de}}\aunote{These authors contributed equally}, Xuan Zhang\aff{1}$\ddag$, \and Olga Shishkina\aff{1}\corresp{\email{olga.shishkina@ds.mpg.de}}}

\affiliation{\aff{1}Max Planck Institute for Dynamics and Self-Organization,\\ Am Fassberg 17, 37077 G\"ottingen, Germany}

\begin{document}

\maketitle

\begin{abstract}
This work addresses the effects of different thermal sidewall boundary conditions on the formation of flow states and heat transport in two- and three-dimensional Rayleigh--B\'enard convection (RBC) by means of direct numerical simulations and steady-state analysis for Rayleigh numbers $Ra$ up to $4\times10^{10}$ and Prandtl numbers $Pr=0.1,1$ and $10$. We show that a linear temperature profile imposed at the conductive sidewall leads to a premature collapse of the single-roll state, whereas a sidewall maintained at a constant temperature enhances its stability. The collapse is caused by accelerated growth of the corner rolls with two distinct growth rate regimes determined by diffusion or convection for small or large $Ra$, respectively. Above the collapse of the single-roll state, we find the emergence of a double-roll state in two-dimensional RBC and a double-toroidal state in three-dimensional cylindrical RBC. These states are most prominent in RBC with conductive sidewalls. The different states are reflected in the global heat transport, so that the different thermal conditions at the sidewall lead to significant differences in the Nusselt number for small to moderate $Ra$. However, for larger $Ra$, heat transport and flow dynamics become increasingly alike for different sidewalls and are almost indistinguishable for $Ra>10^9$. This suggests that the influence of imperfectly insulated sidewalls in RBC experiments is insignificant at very high $Ra$ - provided that the mean sidewall temperature is controlled.
\end{abstract}


\section{Introduction}

Understanding thermally induced convection as it arises in the earth's atmospheric/oceanic circulations and deducing its fundamental aspects from laboratory experiments is an ongoing endeavour which motivated numerous experimental and theoretical studies. In this realm, Rayleigh--B\'enard convection (RBC), i.e. a fluid held between two parallel plates heated from below and cooled from above, is the most thoroughly investigated model system to study the complex physics behind natural convection such as pattern formation and the transition to turbulence \citep{Bodenschatz2000, Ahlers2009, Lohse2010}.

Most of the early theoretical advances were made by considering the system as infinitely extended in the lateral direction. For instance, conventional linear-stability analysis predicts the formation of two-dimensional rolls \citep{Chandrasekhar1961}, while a weakly non-linear analysis reveals the stability regimes of these rolls and their path to subsequent oscillatory or stationary type bifurcations \citep{Schlueter1965, Busse1967, Busse1978}. In laboratory experiments, however, we must resort to laterally confined systems where our understanding is far less complete. In particular, when the lateral size of the container is close to or less than the height of the cell, the presence of sidewalls plays an important role \citep{Roche2020, Shishkina2021}. Therefore, this study focuses on the effects of different thermal sidewall boundary conditions on heat transfer and the emergence of different flow states.

Different sidewalls are known to affect the critical Rayleigh number $Ra_c$ above which convection sets in \citep{Buell1983, Hebert2010}, and perfectly conducting sidewalls have been found to delay the onset compared to adiabatic sidewalls. In an attempt to better understand the flow regimes above onset, bifurcation analyses were performed in a cubic domain for adiabatic \citep{Puigjaner2004} and perfectly conducting sidewalls \citep{Puigjaner2008}. The bifurcation diagrams for the conducting sidewalls are generally more complex, and double-toroidal states predominate over the classical single-roll structure found for adiabatic sidewalls. Sidewalls also have a strong influence on pattern formation \citep{Cross1993, Bruyn1996, Bodenschatz2000} and different sidewall boundary conditions lead to differences in observable patterns even in cells with large aspect ratio \citep{Hu1993}. 

In RBC experiments, spurious sidewall heat fluxes are a major practical difficulty that can substantially bias global heat transport measurements. \cite{Ahlers2000} reported that naive sidewall corrections can overstate Nusselt number measurements by up to $20\%$ and underestimate the scaling of the Nusselt number $Nu$ with respect to the Rayleigh number $Ra$ ($Nu \sim Ra^\lambda$) reflected in the reduction of the scaling exponent $\lambda$ by about $2\%$, underscoring the importance of more sophisticated sidewall corrections. \cite{Roche2001a} further emphasized this conclusion by showing that the sidewall corrections can be considerably larger than assumed, leading to scaling exponents closer to the turbulent scaling of $Nu \sim Ra^{1/3}$ \citep{Grossmann2000, Grossmann2001, Grossmann2004} than previously measured. Probably the most important question in convection today is whether the ultimate regime in confined geometries has the same scaling as predicted for unbounded domains, i.e. $Nu \sim Ra^{1/2}$ (up to different logarithmic corrections), as proposed by \cite{Kraichnan1962} and \cite{Grossmann2011}. Another important question is when and how exactly the transition to the ultimate regime takes place in confined geometries. Laboratory experiments \citep{Chavanne1997, Niemela2000, Chavanne2001, Ahlers2009c, Ahlers2012b, He2012, Urban2014, Roche2020} in this extremely high $Ra$ regime are notoriously difficult to perform and potentially sensitive to several unknowns of the system, one of which is the influence of imperfectly isolated/adiabatic sidewalls.

Numerical simulations were performed incorporating thermal conduction in the solid sidewall to clarify the differences between an ideal adiabatic setup and a finite thermal conductivity sidewall \citep{Verzicco2002, Stevens2014, Wan2019}. The results of these studies suggest that different thermal properties of the sidewall alter the mean flow structure, leading to significant differences in global heat transport in the low to mid $Ra$ range. However, this effect vanishes for larger $Ra$, at least when the sidewall temperature is constant and maintained at the arithmetic mean of upper and lower plate temperatures. Conversely, if the sidewall temperature deviates from the arithmetic mean, differences in heat transport persist even for large $Ra$. This indicates that it is more important to keep the environment at the correct temperature than to shield the interior of the cell from its surroundings.

Despite extensive previous work, the spatial distribution of flow and heat transport in confined geometries with different thermal boundary condition has not been exhausted, especially the conditions related to real experimental sidewall boundary conditions. In the present work, we investigate RBC with the following thermal sidewall boundary conditions: adiabatic, constant temperature (isothermal) and linear temperature. In the first part of the results, we focus on a steady-state analysis based on an adjoint descent algorithm \citep{Farazmand2016} to identify different flow states, their properties and their evolution over $Ra$. In the second part, the analysis is complemented and extended to higher $Ra$ into the turbulent regime by a set of DNS for a 2D box and 3D cylindrical setup, covering a range of $10^3<Ra<10^{11}$ and $10^3<Ra<10^{9}$, respectively, aiming for a more complete picture. We first present our numerical methods, discuss the results and conclude with our main findings. 

\section{Numerical methods}

\subsection{Governing equations}

The dimensionless control parameters in RBC are the Rayleigh number $\Ra \equiv \alpha g \Delta H^3/(\kappa \nu)$,
the Prandtl number $\Pran\equiv \nu/\kappa$, and the width-to-height aspect ratio of the box, $\Gamma\equiv L/H$.
Here, $\alpha$ denotes the isobaric thermal expansion coefficient, $\nu$ the kinematic viscosity, $\kappa$ the thermal diffusivity of the fluid, $g$ the acceleration due to gravity, $\Delta\equiv T_+-T_-$ the difference between the temperatures at the lower ($T_+$) and upper ($T_-$) plates, $H$ the distance between the parallel plates (the container height), and $L$ the length of the container or the diameter  in the case of a cylindrical setup. In this study, we focus on variations with $Ra$, while $Pr=1$ is fixed for most results in this paper except for a $Pr$-dependence study in section \ref{sec:pr-dep}, and $\Gamma=1$ is held constant throughout the study.

The governing equations in the Oberbeck--Boussinessq approximation for the dimensionless, incompressible velocity ${\bf u}$, temperature $\theta$ and kinematic pressure $p$ read as follows: 
\begin{eqnarray}
   \partial{\bf u}/\partial t+{\bf u}\cdot \nab {\bf u}+\nab {p}&=&
    \sqrt{Pr/Ra} \nab^2 {\bf u}+ {\theta}{\bf e}_z , \nonumber\\
    \partial{\theta}/\partial t+{\bf u}\cdot \nab {\theta}&=&1/\sqrt{Pr Ra} \nab^2 {\theta}, \quad \nab \cdot {\bf u} =0.
\label{eq:ns}
\end{eqnarray}
The equations were made dimensionless using the free-fall velocity $u_{ff}\equiv (\alpha g \Delta H)^{1/2}$, the free-fall time $t_{ff}\equiv H/u_{ff}$, the temperature difference $\Delta\equiv T_+-T_-$ between bottom ($T_+$) and top ($T_-$) plates and $H$ the cell height. Here ${\bf e}_z$ is the unit vector in the vertical $z$-direction. This set of equations is solved with the direct numerical solver {\sc goldfish}, which uses a fourth-order finite volume discretization on a staggered grid and a third order Runge--Kutta time scheme. The code has been widely used in previous studies and validated against other direct numerical simulation codes \citep{Kooij2018, Reiter2021a}.

\subsection{Boundary conditions}
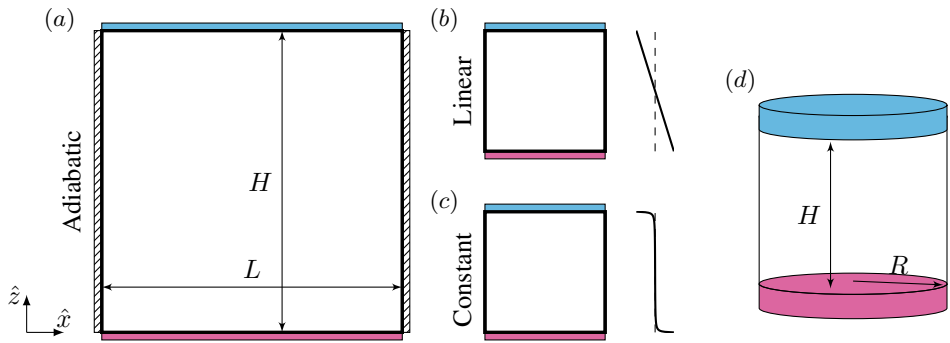
\begin{figure}
\unitlength1truecm
\begin{picture}(12, 5.5)
\put(0.25, 0.5){\includegraphics[width=12.5cm]{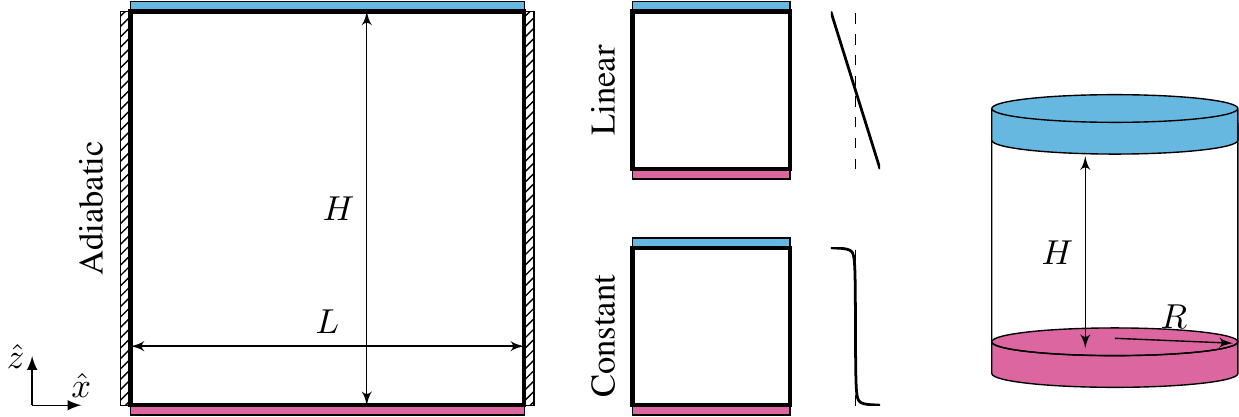}}
\put(0.8, 4.65){$(a)$}
\put(5.9, 4.65){$(b)$}
\put(5.9, 2.25){$(c)$}
\put(9.8, 3.85){$(d)$}
\end{picture}
\caption{
2D Numerical setup of $(a)$ adiabatic, $(b)$ linear and $(c)$ constant sidewall temperature boundary conditions. $(d)$ Sketch of cylindrical domain. Profiles next to $(b)$ and $(c)$ show the imposed sidewall temperature distribution.
}
\label{fig:setup}
\end{figure}

We study 2D RBC in a square box and 3D RBC in a cylindrical domain. The setups and profiles of the sidewall (SW) boundary conditions (BCs) used are shown in figure \ref{fig:setup}. The adiabatic, linear and constant conditions for the sidewall region $\delta V_S$ are defined by
\begin{align}
    \text{adiabatic:}& \quad \partial \theta/\partial{\chi} = 0, \\
    \text{linear:}& \quad \theta = \theta_+ + z \left(\theta_- - \theta_+ \right), \\
    \text{constant:}& \quad \theta =
\begin{cases}
    \frac{-k(2z-1)}{k+2 z} \left(\theta_+ - \theta_m \right),  &0 \leq z\leq 1/2,\\ 
    \frac{k(2z-1)}{k-2z+2}\left(\theta_- - \theta_m \right),  & 1/2 < z \leq 1,
\end{cases}
\label{eq:zero}
\end{align}
with the temperature of the lower plate $\theta_+=1/2$, the temperature of the upper plate $\theta_-=-1/2$, their arithmetic mean $\theta_m=0$, $z \equiv z/H \in [0,1]$ and $\chi=x$ for box and $\chi=r$ for cylinder, respectively. As for the constant temperature conditions, most of the sidewall is kept at a nearly uniform temperature ($\theta_m$), except for the transition regions in the vicinity of the top and bottom plates to ensure a smooth temperature distribution. The parameter $0<k\ll 1$ in eq. \eqref{eq:zero} defines the thickness of the transition layer. Here we used $k=0.01$, which gives a fairly sharp albeit sufficiently smooth transition, as can be seen in figure \ref{fig:setup} $(c)$. Moreover, the velocity no-slip conditions apply to all walls, i.e. $\mathbf{u}\eval_\text{wall}= 0$.

\subsection{Adjoint descent method}
\label{sec:adj_descent}

A complementary analysis to direct numerical simulations is the study of the Boussinesq equations by means of its invariant solutions. \cite{Hopf1948} conjectured that the solution of the Navier--Stokes equations can be understood as a finite but possibly large number of invariant solutions, and turbulence from this point of view is the migration from the neighbourhood of one solution to another. While highly chaotic systems seem hopelessly complex to understand, laminar or weakly chaotic flows can often be captured quite well with this approach. In this work, we focus solely on solutions for steady-states (equilibrium).

Determining steady-state solutions can be quite difficult, especially when the number of dimensions is large as it is the case for most fluid mechanical problems. The most commonly used numerical method for this task is Newton's method, which usually uses the generalized minimal residual (GMRES) algorithm to solve the corresponding systems of linear equations \citep{Saad1986}. This method generally shows fast convergence rates when the initial estimate is close to the equilibrium point. However, if the initial estimate is too far from the equilibrium, Newton's method often fails. In particular, for fluid mechanics, the basin of attraction of Newton's method can be quite small, making the search for steady-states highly dependent on the initial guess. Here we consider an alternative approach recently proposed by \cite{Farazmand2016} based on an adjoint method. \cite{Farazmand2016} has shown that this adjoint-descent method can significantly improve the chance of convergence compared to the Newton--descent method, and thus more reliably capture equilibrium states from a given initial state, but at the cost of a generally slower convergence rate. A detailed derivation of the algorithm can be found in \cite{Farazmand2016}. Below we sketch the idea of the method.

Suppose we want to find equilibrium solutions of a particular PDE (in our case the Boussinessq equations)
\begin{equation}
	\partial_t \uu = F(\uu),
\label{eq:PDEF}
\end{equation}
with $\uu=\uu(\mathbf{x}, t)$. The equilibrium's of F(\uu) can be generally unstable and therefore difficult to detect. The idea is to search a new PDE, i.e.
\begin{equation}
	\partial_\tau \uu = G(\uu),
\label{eq:G}
\end{equation}
which solutions always converge to the equilibrium solutions of \eqref{eq:PDEF} when the fictitious time $\tau$ goes to infinity
\begin{equation}
\norm{F(\uu)}_\mathcal{A}^2 \rightarrow 0 \quad \text{as} \quad \tau \rightarrow \infty,
\label{eq:to_zero}
\end{equation}
with the weighted energy norm $\norm{\cdot}_\mathcal{A} \equiv \langle \cdot, \cdot \rangle_\mathcal{A} \equiv\langle \cdot, \mathcal{A} \cdot \rangle$ for a certain real self-adjoint and positive definite operator $\mathcal{A}$. $F(\uu)$ evolves along a trajectory $\uu^\prime$ in accordance with
\begin{equation}
\frac{1}{2} \partial_\tau \norm{F(\uu)}_\mathcal{A}^2 = \langle \delta F(\uu, \uu^\prime), F(\uu) \rangle_\mathcal{A},
\label{eq:evolve}
\end{equation}
where $\delta F(\uu,\uu^\prime) \equiv \lim \limits_{\varepsilon \to 0}\frac{F(\uu + \varepsilon \uu^\prime)-F(\uu)}{\varepsilon}$ of $F(\uu)$ is the functional Gateaux derivative at $\uu$ in the direction $\uu^\prime$. In the Newton-descent method, the search direction $\uu^\prime$ is approximated from $\delta F(\uu,\uu^\prime) = - F(\uu)$ by using, for example, a GMRES iterative algorithm. For the adjoint-descent method, on the other hand, we rewrite eq. \eqref{eq:evolve} in the form
\begin{equation}
\frac{1}{2} \partial_\tau \norm{F(\uu)}_\mathcal{A}^2 = \langle  \uu^\prime, \delta F^\dagger(\uu, F(\uu))\rangle_\mathcal{A},
\label{eq:evolve_adj}
\end{equation}
where $\delta F^\dagger$ is the adjoint operator of the functional derivative $\delta F$. For $\uu^\prime = -\delta F^\dagger(\uu, F(\uu))$ one guarantees that $\norm{F(\uu)}_\mathcal{A}^2$ decays to zero along the trajectory $\uu^\prime$, since then $\frac{1}{2} \partial_\tau \norm{F(\uu)}_\mathcal{A}^2 = -\norm{\delta F^\dagger(\uu, F(\uu))}_\mathcal{A}^2$. Letting $\uu$ evolve along the adjoint search direction ensures the convergence to an equilibrium, thus we find the desired PDE $G(\uu) \equiv \uu^\prime$, i.e.
\begin{equation}
G(\uu) = -\delta F^\dagger(\uu, F(\uu)).
\label{eq:g2}
\end{equation}
The choice of the norm $\norm{\cdot}_\mathcal{A}$ is important for the algorithm to be numerically stable and is explained in more detail in the appendix. As mentioned, the operator $\mathcal{A}$ should be real-valued, positive-definite and self-adjoint. Following \cite{Farazmand2016}, we use an operator $\mathcal{A}$ that is closely related to the inversed Laplacian, i.e. $\mathcal{A} = (I - \alpha \nab^2)^{-1}$ where $I$ is the identity operator and $\alpha$ is a non-negative scalar parameter. For $\alpha=0$ this norm converges to the $L^2$-norm and for $\alpha>0$ it effectively dampens smaller scales and provides a better numerical stability. 

The linear adjoint equations for the Boussinesq equations \eqref{eq:ns} read
\begin{align}
    -\partial_\tau\uu &=   \left(\nab \tilde{\uu}\dprime + (\nab \tilde{\uu}^{\prime\prime})^\text{T} \right) \uu + \theta\nab \tilde{\theta}\dprime
    - \nab p\dprime + \sqrt{Pr/Ra} \nab^2 \tilde{\uu}\dprime, \nonumber\\
    -\partial_\tau\theta &= \uu \cdot \nab \tilde{\theta}\dprime + 1/\sqrt{PrRa} \nab^2 \tilde{\theta}\dprime +  \vec{\mathbf{e}}_z\cdot \tilde{\uu}\dprime,\nonumber\\
    \nab \cdot \uu\dprime &=0, \quad \nab \cdot \uu =0
\label{eq:app_adj_eq}
\end{align}
(see derivations in the appendix). Here the double prime fields $\uu\dprime$ and $\theta\dprime$ denote the residuals of the Navier--Stokes eq. \eqref{eq:ns}, i.e.
\begin{align}
    \uu\dprime &\equiv  -\uu \cdot \nab \uu - \nab p + \sqrt{Pr/Ra} \nab^2 \uu + \vec{\mathbf{e}}_z \theta, \nonumber\\
    \theta\dprime &\equiv -\uu \cdot \nab \theta + 1/\sqrt{PrRa} \nab^2 \theta.
\label{eq:residual}
\end{align}
and $\tilde{\uu}\dprime \equiv \mathcal{A}\mathbf{\uu}\dprime$ as well as $\tilde{\theta}\dprime \equiv \mathcal{A}\mathbf{\theta}\dprime$.
For simplicity, let $\mathbf{q} \equiv (\uu, \theta)$, then the adjoint descent method consists of three steps
\begin{enumerate}[leftmargin=1cm, labelsep=0.2cm]
  \item  Find the residuals $\mathbf{q}\dprime$ according to eq. \eqref{eq:residual}.
  \item  Solve $\tilde{\mathbf{q}}\dprime = \mathcal{A}\mathbf{q}\dprime$ for $\tilde{\mathbf{q}}\dprime$.
  \item  Update $\mathbf{q}$ according to eq. \eqref{eq:app_adj_eq}.
\end{enumerate}
In step (i), we solve the time-stepping eq. \eqref{eq:ns}, where we use a standard pressure projection method and treat the diffusion term implicitly. The time step size $\Delta t$ can be chosen independently of the artificial time step size $\Delta \tau$ of the adjoint equations. For step (ii), using the energy norm $\norm{\cdot}_\mathcal{A}$ with the operator $\mathcal{A} = (I - \alpha \nab^2)^{-1}$, we solve the Helmholtz-type equation $(I - \alpha \nab^2)\tilde{\mathbf{q}}\dprime = \mathbf{q}\dprime$. The integration of the adjoint equations in step (iii) is similar to step (i), but all terms are treated explicitly. Through tests, we found that the artificial time step $\Delta \tau$ can be chosen much larger than $\Delta t$ in some cases, i.e. for large $Ra$.

The boundary conditions of $\tilde{\uu}\dprime$ and $\tilde{\theta}\dprime$ result from integration by parts in the derivation of the adjoint equations. Evaluation of the adjoint operator of the diffusion terms yields
\begin{align}
\int_V \tilde{\uu}\dprime \nab^2 \uu^\prime dV = \int_V \uu^\prime \nab^2 \tilde{\uu}\dprime dV + \int_S \uu^\prime(\nab\tilde{\uu}\dprime \cdot \mathbf{n})dS - \int_S \tilde{\uu}\dprime(\nab\uu^\prime \cdot \mathbf{n})dS,
\label{eq:bc}
\end{align}
where we see the occurrence of two additional boundary terms (the last two terms) evaluated on the boundary domain $S$. The first boundary term vanishes since the search direction $\uu^\prime$ is zero on the boundaries. The second term can be eliminated if we also choose homogeneous Dirichlet boundary conditions for the adjoint field $\tilde{\uu}\dprime$ on $S$. The same logic applies to homogeneous Neumann conditions. For the pressure field $p\dprime$, we apply Neumann boundary conditions conditions on all walls.  In this study, all flow states showed good overall convergence ($\norm{F(\uu)}_\mathcal{A}^2\leq 10^{-5}$) and the velocity fields where almost divergence free ($\norm{\div{\uu}}_{L^2}\leq 10^{-3}$). However, the rigorous verification of the chosen pressure BCs has yet to be performed. Another interesting point, reserved for later investigation, is whether a vorticity-streamfunction formulation might be better suited to resolve issues with the boundary conditions.

\begin{figure}
\unitlength1truecm
\begin{picture}(12, 5.5)
\put(3.0, 0.0){\includegraphics[width=7.0cm]{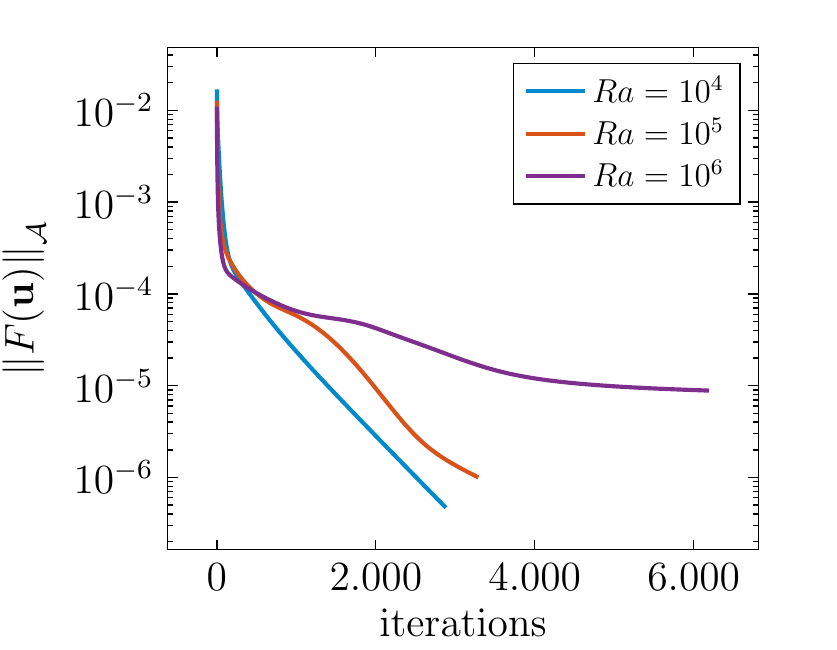}}
\end{picture}
\caption{Convergence of the adjoint-descent method for three different $Ra$, starting from the same initial field. The time-step size for which the algorithm is just stable increased with $Ra$, i.e., for these cases we used $\Delta \tau = 0.5$ ($Ra=10^4$),  $\Delta \tau = 2.0$ ($Ra=10^5$) and $\Delta \tau = 5.0$ ($Ra=10^6$). All three cases converged to large-scale circulation flow states as described in section \ref{sec:lsc}.
}
\label{fig:steady_convergence}
\end{figure}

For the steady-state analysis, we use a Galerkin method with Chebyshev bases in $x$ and $z$ directions and a quasi-inverse matrix diagonalization strategy for better efficiency \citep{Shen1995, Julien2009, Oh2019, Mortensen2018}. The code is publicly available \citep{Reiter2021b}. We use an implicit backward Euler time discretization and alias the fields using the $2/3$ rule by setting the last $1/3$ high-frequency spectral coefficients to zero after evaluating the nonlinear terms. When used as a direct numerical solver, we found excellent agreement with our finite-volume code {\sc goldfish}. In addition, the steady-states from the adjoint descent method showed excellent agreement with those found by an alternative Newton--GMRES iteration. Figure \ref{fig:steady_convergence} shows the convergence rates for three different $Ra$, starting from the same initial state. Overall, we find that the convergence chance is improved over the Newton-descent method, although the convergence rate suffers and larger $Ra$ are either not feasible with the current approach as implemented in our code or diverge after some time. Therefore, we restrict the steady-state analysis to flows in the range $Ra\leq10^7$ and investigate larger $Ra$ using direct numerical simulations. One conceivable problem with the current approach is that the currently used energy norm with the operator $\mathcal{A}\equiv (I - \alpha \nab^2)^{-1}$ dampens smaller scales in order to increase the stability of the algorithm. But for larger $Ra$, smaller scales become important to resolve the boundary layers sufficiently, so the algorithm is likely to take longer to converge or the damping of the smaller scales is too severe to reach convergence overall. Using smaller values of $\alpha$ could lead to better results in that case, as it emphasizes smaller scales more. Preliminary analysis suggests that $\alpha =10^{-3}$ leads to better convergence to a steady-state than $\alpha=1$, but requires smaller time steps $\delta \tau$, which currently makes it too costly to apply to a wider range of parameters. In the future, the convergence rate might be improved by employing a hybrid adjoint-descent and Newton-GMRES approach, as proposed by \cite{Farazmand2016}. Alternative gradient optimization techniques are also conceivable to boost convergence speed.

\section{Steady-state analysis}

\begin{figure}
\unitlength1truecm
\begin{picture}(12, 5.0)
\put(0.0, 0.0){\includegraphics[height=4.6cm]{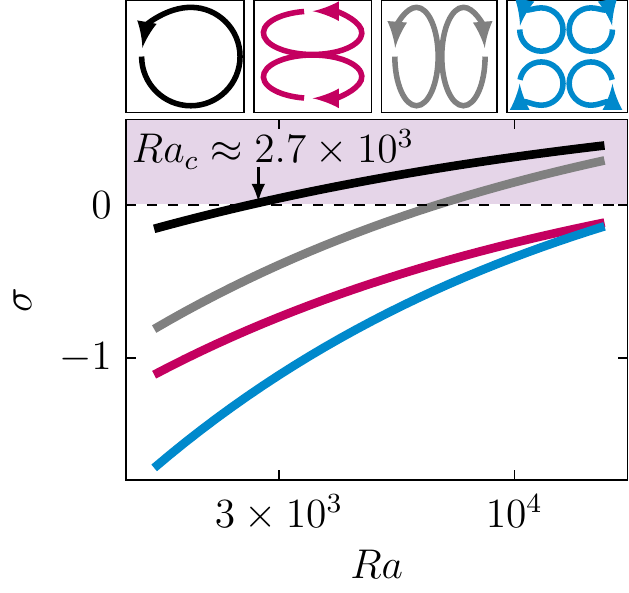}}
\put(5.0, 0.0){\includegraphics[height=4.6cm]{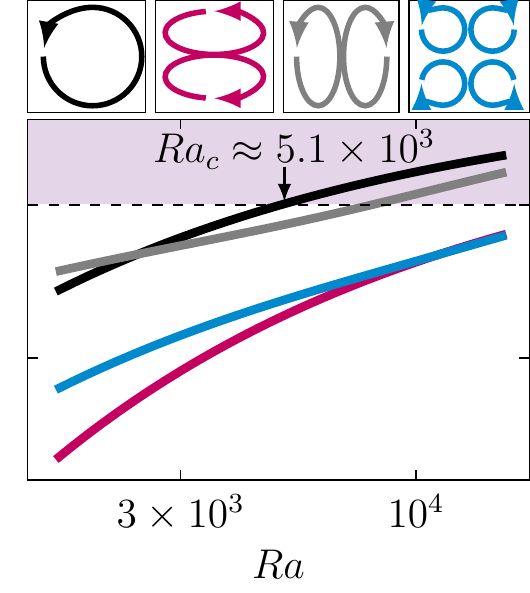}}
\put(9.25, 0.0){\includegraphics[height=4.6cm]{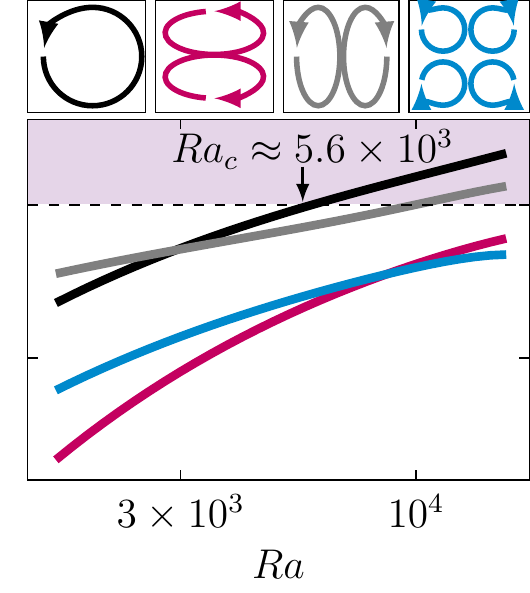}}
\put(0.5, 4.85){$(a)$}
\put(4.9, 4.85){$(b)$}
\put(9.1, 4.85){$(c)$}
\put(1.8, 4.85){Adiabatic SW}
\put(6.4, 4.85){Linear SW}
\put(10.5, 4.85){Constant SW}
\end{picture}
\caption{Growth rates $\sigma$ as determined from linear stability analysis for the four most unstable modes at the onset of convection in the 2D cell for $(a)$ adiabatic, $(b)$ linear and $(c)$ constant sidewall boundary conditions. The most unstable modes are schematically depicted above each graph with the corresponding colour. The critical Rayleigh numbers for the current convection, $Ra_c$, are marked with errors.}
\label{fig:onset}
\end{figure}
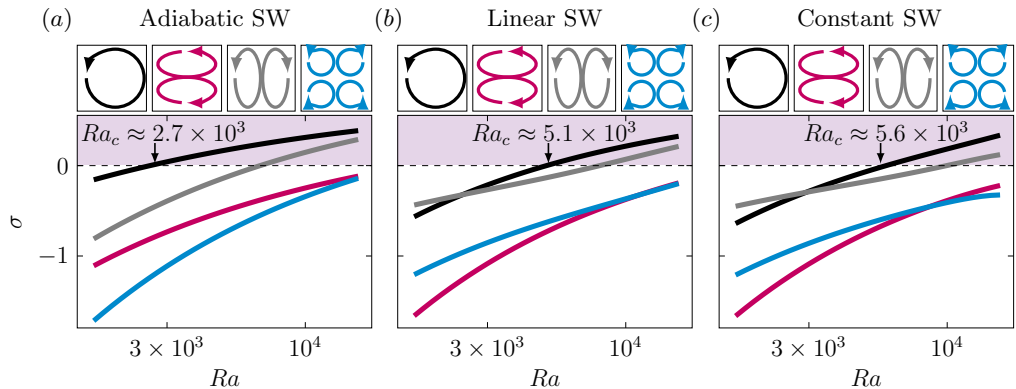

In this section, we study steady-states in 2D RBC for $Ra\leq10^7$. In what follows, we refer to flow states as single or multiple solutions connected by inherent symmetries of the system. For example, the single-roll state (SRS) in 2D can exist in two forms, either circulating clockwise or counterclockwise, but is considered as a single flow state that is invariant under reflection. Steady-state solutions of the SRS state have been investigated in laterally periodic flows with stress-free velocity boundary conditions on the horizontal walls \citep{Wen2015, Wen2020} and with no-slip BCs \citep{Waleffe2015, Sondak2015, Wen2020a, Kooloth2021}. Bifurcations and different flow states have already been studied in laterally unbounded RBC \citep{Zienicke1998}, in laterally bounded RBC for a cubic domain \citep{Puigjaner2008} and a 2D square domain \citep{Venturi2010}. Here we focus on the onset of convection, the SRS and a vertically stacked double-roll state (DRS) in two-dimensional RBC for three different sidewall BCs as shown in figure \ref{fig:setup}.

\subsection{Onset of Convection}

In RBC, there is a critical Rayleigh number $Ra_c$ above which the system bifurcates from the conduction state to coherent rolls. We calculate $Ra_c$ using a linear stability analysis described in more detail in \cite{Reiter2021}. For adiabatic or linear (conductive) sidewall BCs, the conduction or base state is characterized by a linear temperature profile in the vertical direction with zero velocity field and independence from control parameters. However, for a constant temperature sidewall distribution, a convective flow is already present. In this case, we perform a steady-state search before analyzing the local stability around this equilibrium point.

Figure \ref{fig:onset} shows the linear growth rates of the four most unstable modes, which resemble the first four Fourier modes as depicted in the same figure. All three BCs initially bifurcate from the conduction state to a single roll state. Adiabatic sidewalls lead to a lower critical Rayleigh number compared to isothermal sidewalls, which is to be expected \citep{Buell1983}. The onset for the adiabatic sidewall occurs at $Ra_c \approx 2.7 \times 10^3$ which agrees well within our resolution limit with \cite{Venturi2010}, who reports a critical $Ra$ of about $2582$. The onset for the linear SW occurs at $5.1\times 10^3$ and the onset for the constant SW occurs slightly later at $5.6\times 10^3$. This indicates that the interaction of the convective field - as present for the constant sidewall BC - with the unstable modes is weak and its influence on the onset is small.

\subsection{Single-roll (states $\mathcal{S}_A^1$, $\mathcal{S}_L^1$, $\mathcal{S}_C^1$)}
\label{sec:lsc}

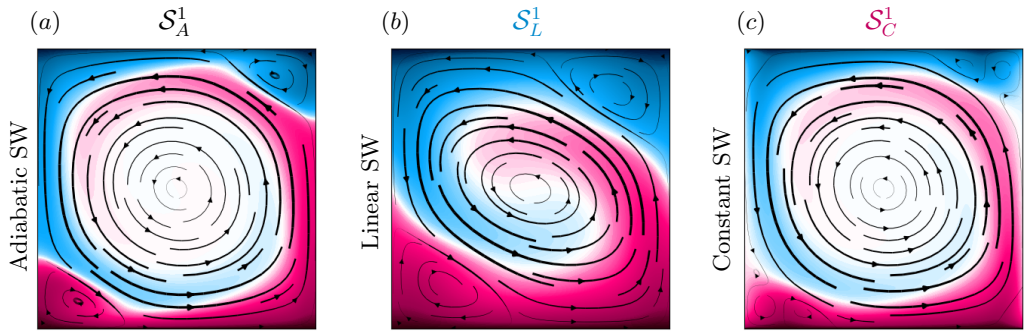
\begin{figure}
\unitlength1truecm
\begin{picture}(12, 4.5)
\put(0.1, 0.0){\includegraphics[height=4.6cm]{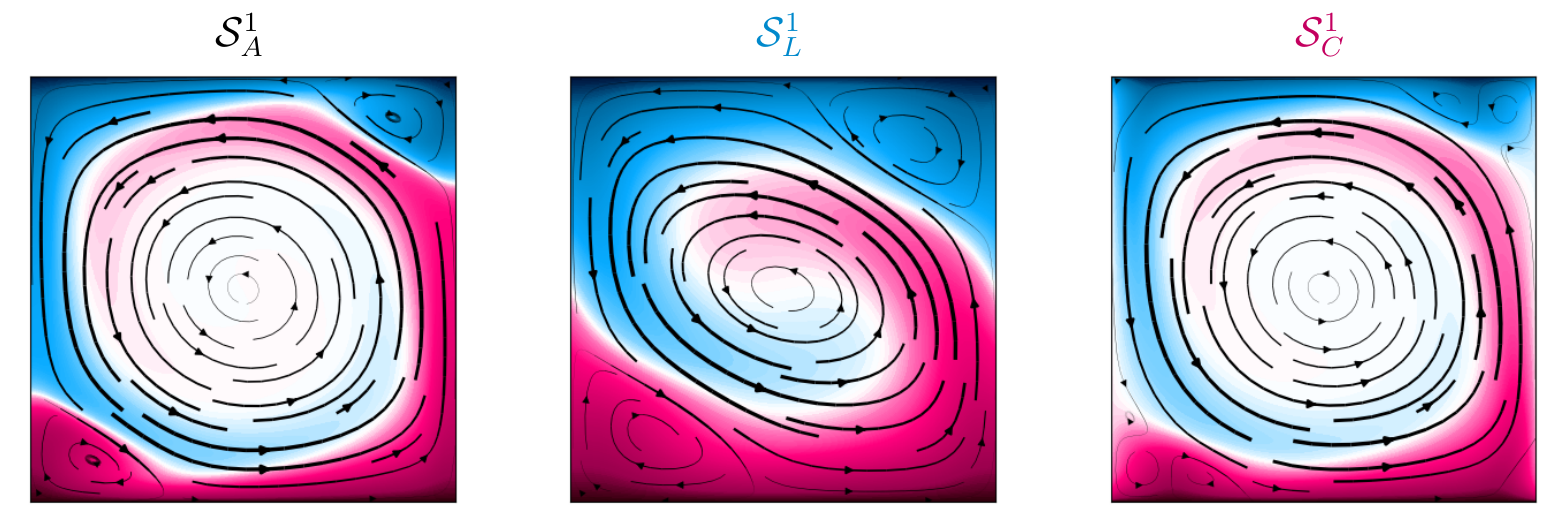}}
\put(0.25, 4.2){$(a)$}
\put(0,1){\rotatebox{90}{Adiabatic SW}}
\put(4.9, 4.2){$(b)$}
\put(4.65,1.3){\rotatebox{90}{Linear SW}}
\put(9.6, 4.2){$(c)$}
\put(9.3,1){\rotatebox{90}{Constant SW}}
\end{picture}
\caption{
Single roll state for $(a)$ adiabatic ($Ra=10^6$), $(b)$ linear ($Ra=9\times 10^4$) and $(c)$ constant ($Ra=10^6$) sidewall temperature boundary conditions. Contours (streamlines) represent the temperature (velocity) field.
}
\label{fig:steady_fields_roll1}
\end{figure}

The single roll state (SRS) is arguably the most important state in RBC. It is the first mode to appear above the conduction state, as we have just seen, and prevails even up to largest $Ra$ in the form of large-scale circulation (LSC) on turbulent superstructures \citep{Zhu2018, Reiter2021a}. The SRS is stable and time-independent for small $Ra$ but oscillatory, chaotic, or even completely vanishing for larger $Ra$, as we will show in section \ref{sec:modal}. Here we analyze its properties before collapse and show that the growth of secondary corner rolls plays an important role in its destabilization and that this process can be both suppressed and enhanced by different sidewall boundary conditions.

Figure \ref{fig:steady_fields_roll1} shows the temperature and velocity fields of the SRS for different sidewall BCs. For all three BCs we can identify a large primary roll circulating counter-clockwise and two secondary corner rolls. The corner rolls are most pronounced for the linear sidewall BC and the primary roll is nearly elliptical. The dimensionless heat-flux is expressed in form of the Nusselt number $Nu\equiv \sqrt{RaPr} F_f H / \Delta$ with the heat-flux $F_f$ entering the fluid and the imposed temperature difference $\Delta$. $F_f$ can be defined in different ways, especially in the presence of sidewall heat-fluxes. Averaging the temperature equation in eq. \eqref{eq:ns} over time, one obtains
\begin{align}
\div \mathbf{F}=0, \quad \mathbf{F}\equiv \uu \theta -1/\sqrt{RaPr} \nab{\theta},
\label{eq:F}
\end{align}
from which it follows that the total heat flux must vanish through the boundaries $S=\delta V$, i.e. $\int_S (F\cdot \mathbf{n}) dS = 0$. For isothermal sidewall BCs, asymmetric flow states with net nonzero sidewall heat-fluxes are possible; in this case the heat fluxes through the bottom and top plates would deviate from each other. However, in the present study, we found that all sidewall heat fluxes are approximately equal to zero when integrated vertically and the temperature gradient at the bottom plate is approximately equal to the temperature gradient at the top plate. Therefore, we define $Nu$ based on the lower (hot) plate at $z=0$:
\begin{equation}
    Nu \equiv  -\frac{1}{A_+}\int_{S_+} \frac{\partial \theta}{\partial z}  d{S_+},
    \label{eq:Nu}
\end{equation}
with the bottom plate domain $S_+$ and its surface area $A_+$. The dimensionless momentum transport is given by the Reynolds number
\begin{equation}
    Re \equiv  \sqrt{Ra/Pr} \sqrt{\langle \mathbf{U}^2\rangle_V} L ,
    \label{eq:Re}
\end{equation}
based on total kinetic energy of the mean field velocity $\mathbf{U}$. Here, $\langle \cdot \rangle_V$ denotes a volume average.

\begin{figure}
\unitlength1truecm
\begin{picture}(12, 4.5)
\put(0.0, 0.0){\includegraphics[height=3.9cm]{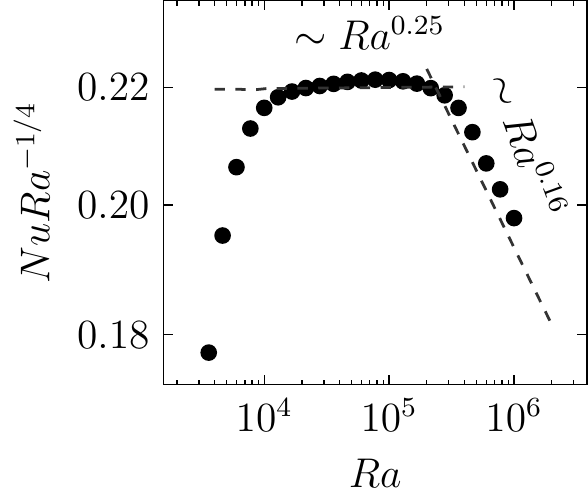}}
\put(4.75, 0.0){\includegraphics[height=3.9cm]{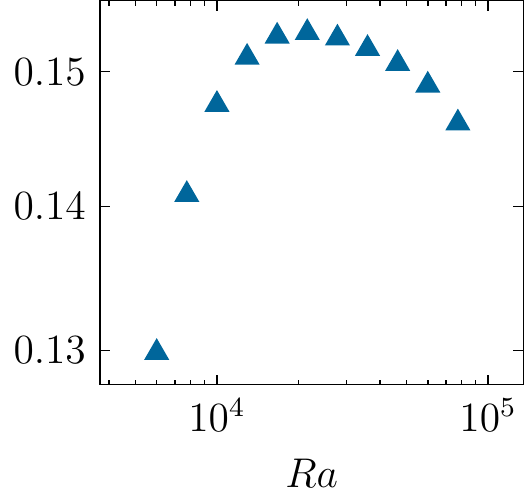}}
\put(9.0, 0.0){\includegraphics[height=3.9cm]{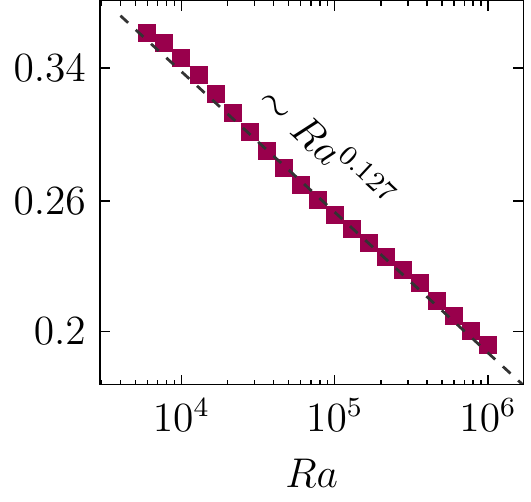}}
\put(0.5, 4.2){$(a)$}
\put(4.8, 4.2){$(b)$}
\put(9.0, 4.2){$(c)$}
\put(1.9, 4.1){Adiabatic SW}
\put(6.4, 4.1){Linear SW}
\put(10.5, 4.1){Constant SW}
\end{picture}
\caption{
Nusselt number $Nu$ for the single-roll states for $(a)$ adiabatic, $(b)$ linear and $(c)$ constant sidewall temperature boundary conditions.
}
\label{fig:roll1_nu}
\end{figure}
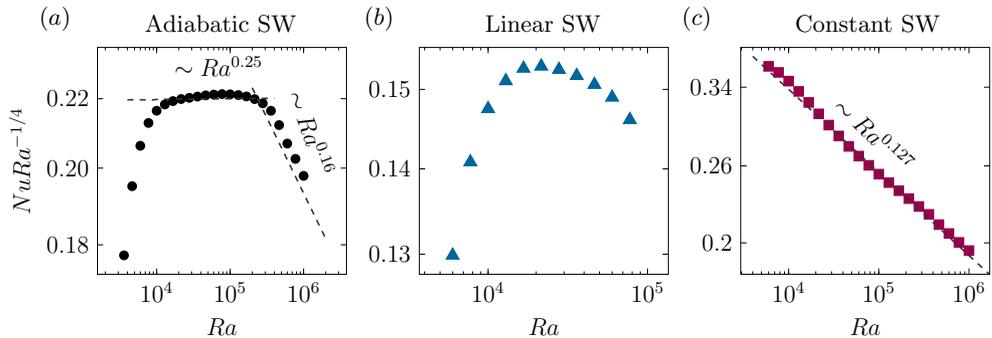

\begin{figure}
\unitlength1truecm
\begin{picture}(12, 4.5)
\put(0.1, 0.0){\includegraphics[height=3.9cm]{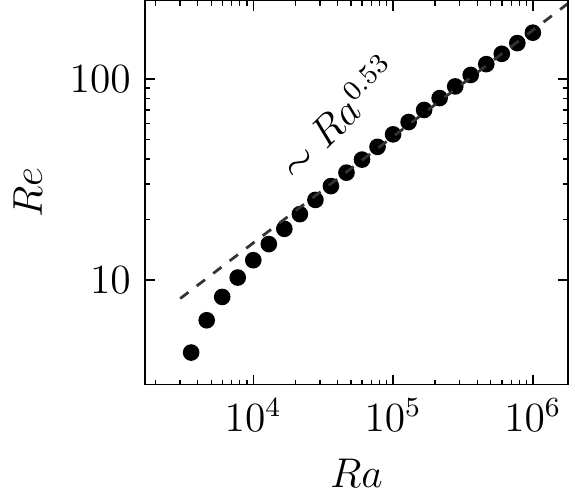}}
\put(4.9, 0.0){\includegraphics[height=3.9cm]{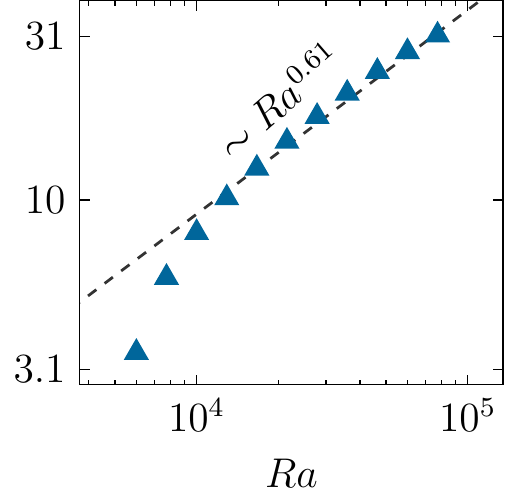}}
\put(9.0, 0.0){\includegraphics[height=3.9cm]{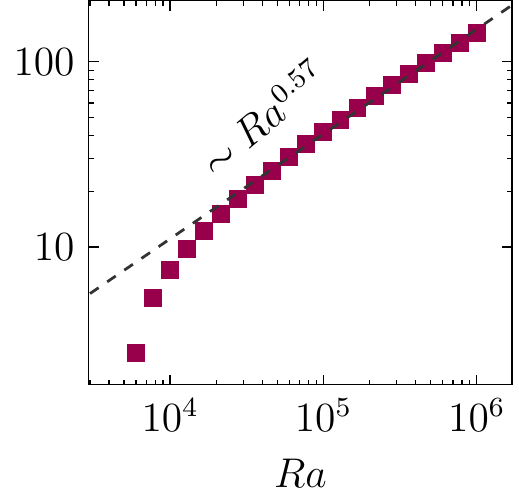}}
\put(0.5, 4.2){$(a)$}
\put(4.8, 4.2){$(b)$}
\put(9.0, 4.2){$(c)$}
\put(1.9, 4.1){Adiabatic SW}
\put(6.4, 4.1){Linear SW}
\put(10.5, 4.1){Constant SW}
\end{picture}
\caption{
Reynolds number $Re$ for the single roll states $\mathcal{S}_A^1$, $\mathcal{S}_L^1$ , $\mathcal{S}_C^1$. $(a)$ adiabatic, $(b)$ linear and $(c)$ constant sidewall temperature boundary conditions.
}
\label{fig:roll1_re}
\end{figure}
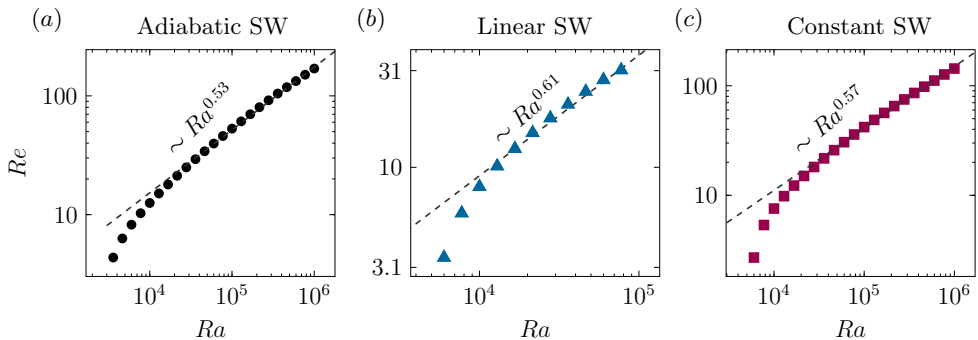

In the laminar regime, where the dissipation of velocity and temperature field is determined by the contributions of the boundary layers, we expect the total heat and momentum scaling $Nu \sim Ra^{1/4}$ and $Re \sim Ra^{1/2}$ \citep{Grossmann2000}, respectively. Figure \ref{fig:roll1_nu} shows that the former scaling shows up only for a very limited $Ra$ range and only for the adiabatic boundary conditions. The SRS of the linear sidewall BCs is stable only up to $Ra\leq 10^5$, then the corner rolls become strong enough to lead to a collapse of the SRS. The stability region where the steady-states converge is too small to observe an unperturbed scaling. On the other hand, for the constant sidewall boundary conditions, corner roll growth is less dominant. In this case, the reason why $Nu$ scaling deviates from $1/4$, is that heat entering through the bottom/top can immediately escape through the sidewalls in the form of a "short-circuit", which dominates the lower $Ra$ regime and is the reason why $Nu$ is relatively large for small $Ra$. For the adiabatic sidewall BC, we observe $Nu \sim Ra^{0.25}$ for $10^4\leq Ra \leq 3\times 10^5$, followed by $Nu\sim Ra^{0.16}$ for $3\times 10^5\leq Ra\leq 10^6$. Similarly, the growth of the corner rolls disturbs the convection wind, and $Nu$ deviates from the ideal $1/4$ scaling. Looking at the $Re$ vs. $Ra$ scaling in figure $\ref{fig:roll1_re}$, we find the theoretically predicted scaling of $1/2$ is better represented in comparison and the different sidewall boundary conditions deviate less among themselves. This suggests that momentum transport is less affected by changing sidewall boundary conditions than heat transport.

\subsubsection{Growth of corner rolls}
The SRS is stable up to a certain $Ra$ limit. Above this limit, it may fluctuate, reverse orientation, or even disappear altogether. This process occurs at $Ra\approx 10^6$ for the adiabatic and constant temperature sidewall BCs and at $Ra\approx 10^5$ for the linear sidewall BC. While up to this event the dynamic behaviour of the three different sidewall BCs is qualitatively very similar, from there on it differs. The constant sidewall BC case shows a time dependence, but remains in the SRS state without changing its orientation. The adiabatic and linear sidewall BCs, on the other hand, enter a more chaotic regime of regular and chaotic flow reversals \citep{Xi2007, Sugiyama2010}, some of which are discussed in section \ref{sec:roll2}. Of greatest importance here appears to be the presence and magnification of secondary corner rolls (CRs).

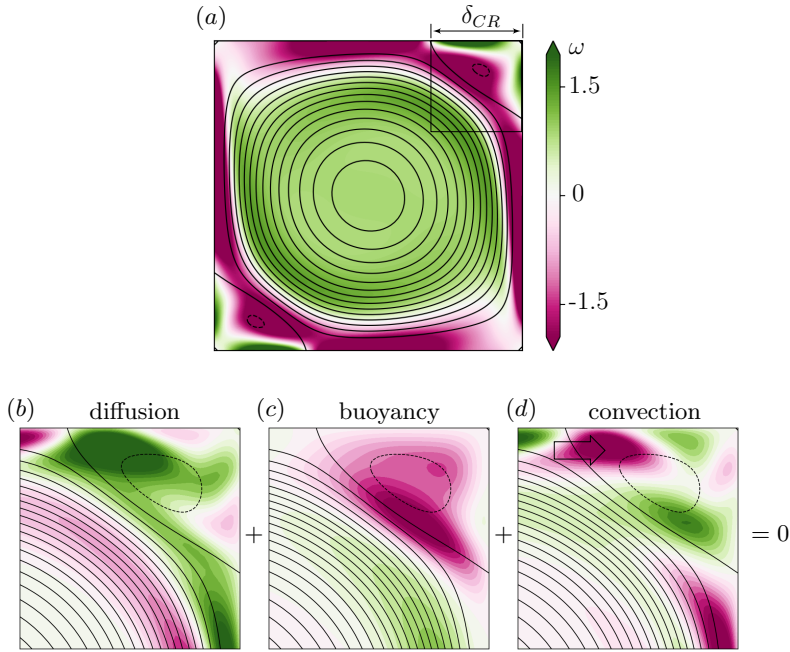
\begin{figure}
\unitlength1truecm
\begin{picture}(12, 9.0)
\put(1.5, 0.0){\includegraphics[width=10.0cm]{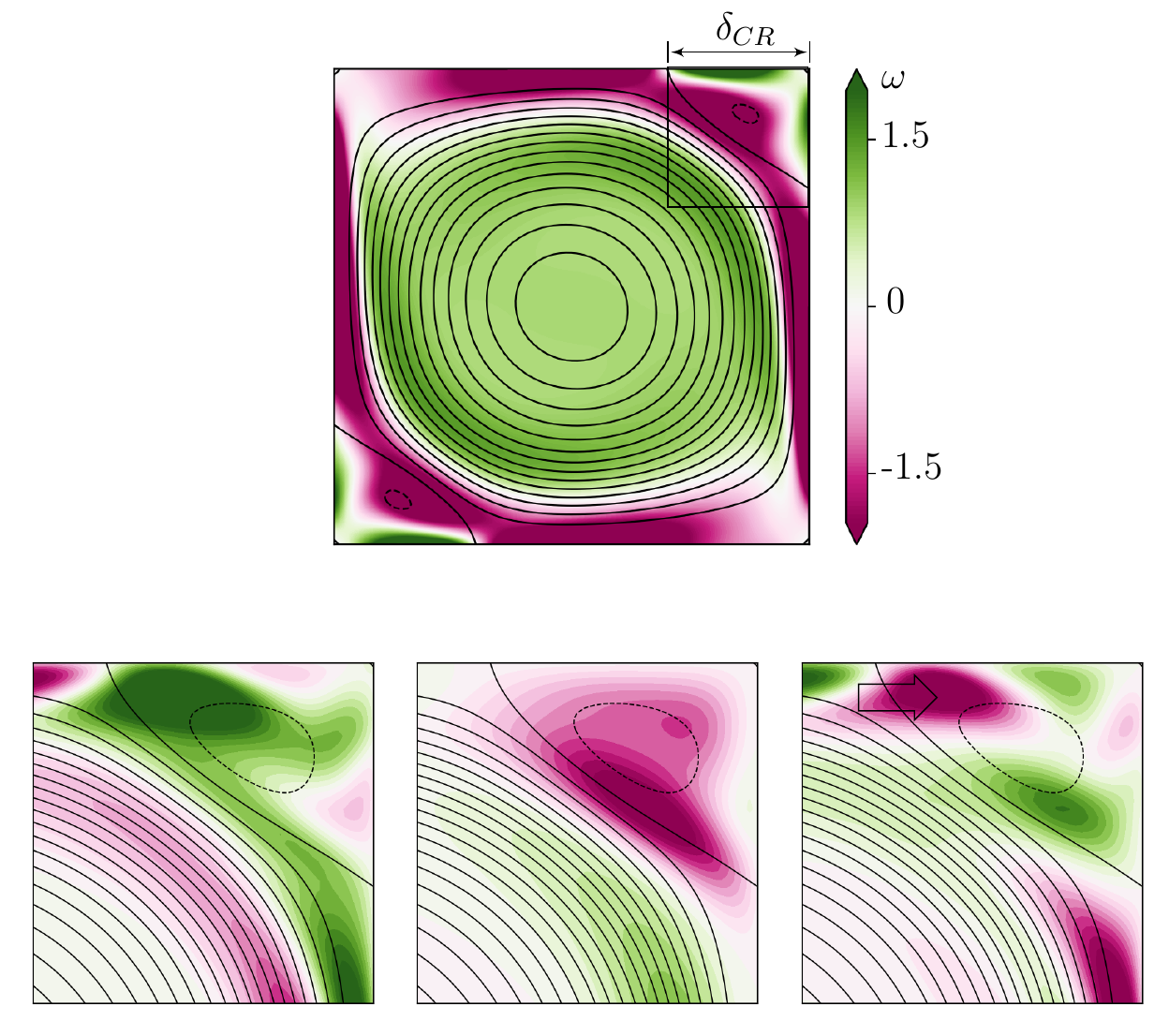}}
\put(4.1, 8.5){$(a)$}
\put(1.6, 3.35){$(b)$}
\put(2.7, 3.3){diffusion}
\put(4.75, 1.7){$+$}
\put(4.9, 3.35){$(c)$}
\put(6.0, 3.3){buoyancy}
\put(8.05, 1.7){$+$}
\put(11.45, 1.7){$=0$}
\put(8.2, 3.35){$(d)$}
\put(9.3, 3.3){convection}
\end{picture}
\caption{
$(a)$ Steady-state vorticity field, velocity streamlines and corner roll size $\delta_{CR}$ defined as a distance from the corner to the closest stagnation point at the plate for $Ra=7\times 10^5$ and adiabatic sidewalls, and vorticity balance contributions according to eq. \eqref{eq:vort} in the corner roll domain, i.e., $(b)$ diffusion, $(c)$ buoyancy and $(d)$ convection. The same contour levers were used for $(b-d)$.
}
\label{fig:vorticity_fields}
\end{figure}

Figure \ref{fig:vorticity_fields} $(a)$ shows the vorticity field and stream-function contour of two-dimensional RBC with adiabatic sidewalls at $Ra=7\times 10^5$. The existence of two corner vortices is apparent. Here we define their size $\delta_{CR}$ based on the zero crossing, or stagnation point, of the vorticity $\omega \equiv \partial_x u_z - \partial_z u_x$ at the top plate, cf. \cite{Shishkina2014}. To understand the processes involved in the formation of the corner rolls, we write down the evolution equation for vorticity
\begin{equation}
    \partial_t \omega = \underbrace{- \uu \cdot \nab \omega}_{\text{convection}} + \underbrace{\sqrt{Pr/Ra} \nab^2 \omega}_{\text{diffusion}} + \underbrace{\partial_x \theta}_{\text{buoyancy}}.
    \label{eq:vort}
\end{equation}
It is evident that for steady-states ($\partial_t \omega = 0$) there must be an equilibrium between convection, diffusion and buoyancy forces. The three corresponding fields are shown in figure \ref{fig:vorticity_fields} $(b-d)$ zoomed in on the corner roll region. For this particular $Ra$, all three contributions appear to be significant. We evaluate the size of the corner rolls (figure \ref{fig:cr_size}) and analyse contributions of diffusion, buoyancy, and convection for all $Ra$ (figure \ref{fig:vorticity_fields}). For this purpose, we evaluate the absolute values of the volume averages for each term in the corner roll region, e.g., $\langle \vert \partial_x \theta \vert \rangle_{V_{CR}}$ represents the strength of the buoyancy term in the corner roll volume $V_{CR}$, as shown in figure \ref{fig:vorticity_fields} $(c)$. The constant BC yields a notable exception because multiple corner rolls can exist. This can be sensed from figure \ref{fig:steady_fields_roll1} $(c)$. For small $Ra$, the corner roll are dominant in the lower right and upper left corner, where the LSC detaches (ejects). For the other two BCs, these rolls are not present. Looking at eq. \eqref{eq:vort}, we realize that the presence of a horizontal temperature gradient can lead to the formation of vortex structures. This condition is present for the constant BCs, e.g., in the lower right corner, where the hot LSC detaches while the temperature is kept constant at zero, resulting in a (strong) negative temperature gradient. The two more "classical" corner rolls first appear at larger $Ra$, but soon take over in size, as can be seen in figure \ref{fig:cr_size}. 

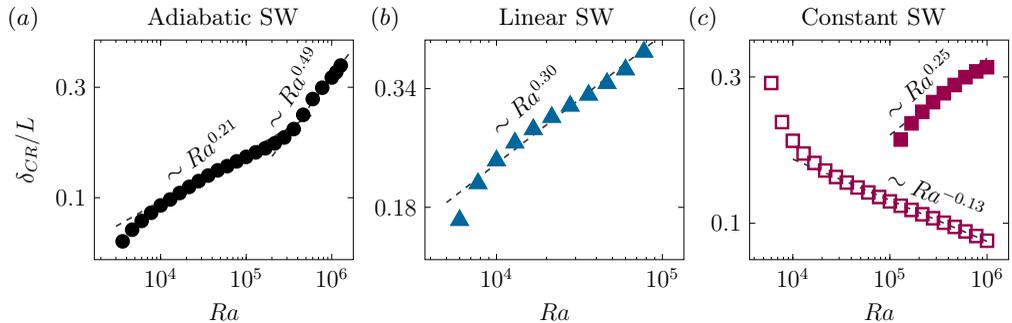
\begin{figure}
\unitlength1truecm
\begin{picture}(12, 4.5)
\put(0.0, 0.0){\includegraphics[height=3.8cm]{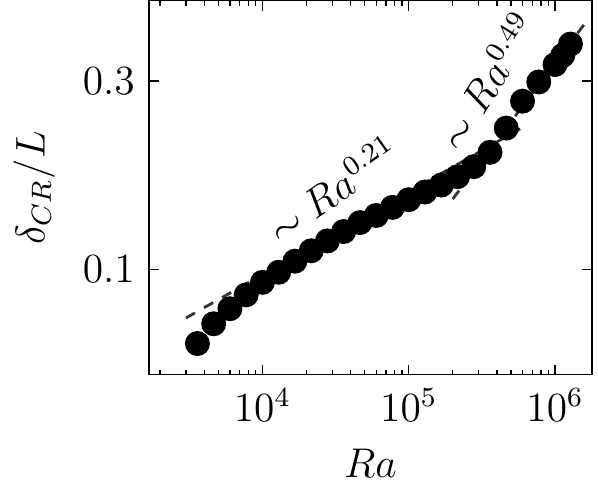}}
\put(4.75, 0.0){\includegraphics[height=3.8cm]{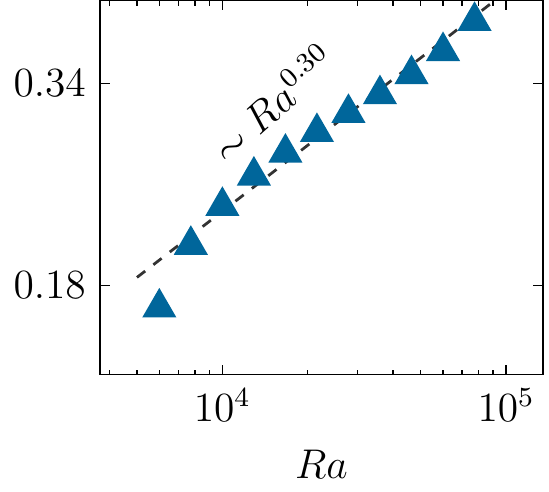}}
\put(9.2, 0.0){\includegraphics[height=3.8cm]{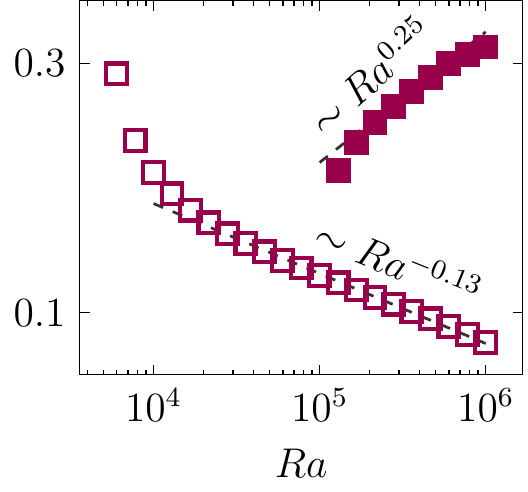}}
\put(0.0, 4.0){$(a)$}
\put(4.8, 4.0){$(b)$}
\put(9.0, 4.0){$(c)$}
\put(1.85, 4.0){Adiabatic SW}
\put(6.5, 4.0){Linear SW}
\put(10.5, 4.0){Constant SW}
\end{picture}
\caption{
Growth of the corner roll size $\delta_{CR}$ for $(a)$ adiabatic, $(b)$ linear and $(c)$ constant sidewall temperature boundary conditions. Adiabatic BC show two distinct regions, a buoyant dominated regime and a regime where convective influx leads to a more rapid increase. For the constant BC, the corner rolls appear first in the plume ejecting corner (bottom right and upper left in figure \ref{fig:steady_fields_roll1}) which is represented by the open symbols in $(c)$, and only for larger $Ra$ do they appear in the plume impacting region (closed symbols).
}
\label{fig:cr_size}
\end{figure}
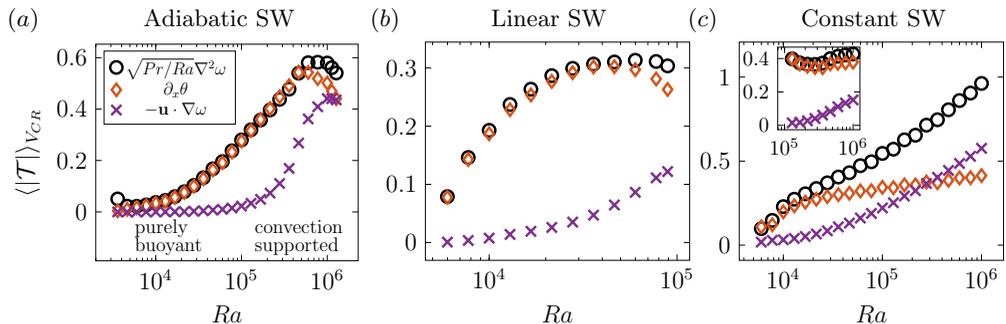
\begin{figure}
\unitlength1truecm
\begin{picture}(12, 4.5)
\put(0.0, 0.0){\includegraphics[height=3.8cm]{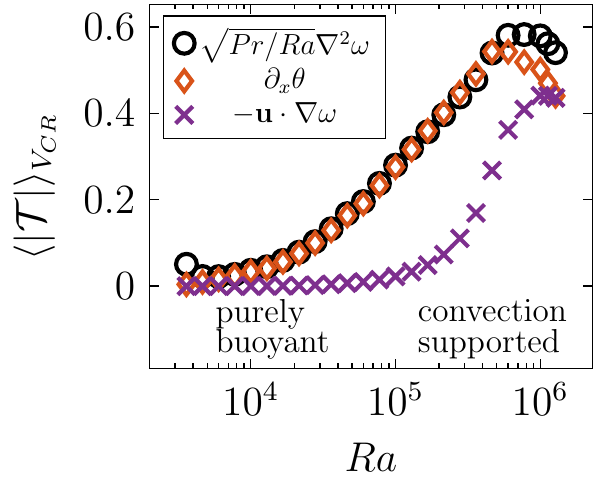}}
\put(4.9, 0.0){\includegraphics[height=3.8cm]{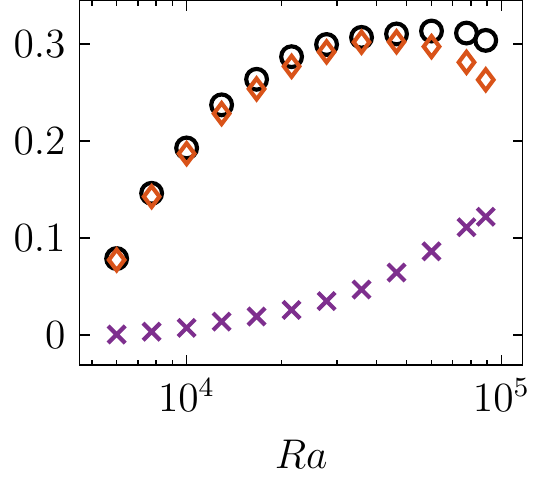}}
\put(9.05, 0.0){\includegraphics[height=3.8cm]{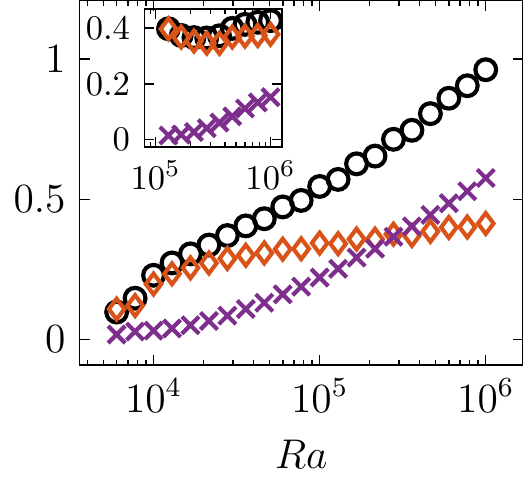}}
\put(0.0, 4.0){$(a)$}
\put(4.8, 4.0){$(b)$}
\put(9.0, 4.0){$(c)$}
\put(1.8, 4.0){Adiabatic SW}
\put(6.4, 4.0){Linear SW}
\put(10.5, 4.0){Constant SW}
\end{picture}
\caption{
Strength of the vorticity balance contributions diffusion (black circles), buoyancy (orange diamonds) and convection (purple pluses) in the corner roll region, according to eq. \eqref{eq:vort}. $(a)$ adiabatic, $(b)$ linear and $(c)$ constant sidewall temperature boundary conditions. Adiabatic BC show two distinct regions, a buoyancy dominated regime and a regime where convective influx leads to a more rapid increase. For the constant BC, the corner rolls appear first in the plume ejecting corner (main figure $c$) and only for larger $Ra$ do they appear in the plume impacting region (inset $c$).
}
\label{fig:vorticity_terms}
\end{figure}
The adiabatic and linear sidewall BCs each yield only two corner rolls. These are present from the onset of convection and grow until the collapse of the SRS (figure \ref{fig:cr_size}). The main difference between the two is that  for the adiabatic sidewall, the corner rolls initially grow monotonically with respect to $Ra$, whereas for the linear sidewall BCs, the corner rolls are already considerable large as soon as the SRS is present. Moreover, they also grow faster with respect to $Ra$ ($\delta_{CR}\sim Ra^{0.3}$) and soon cover almost $40\%$ of the width of the cell. Their large initial size combined with faster growth is the reason for premature SRS instability in linear sidewall BCs. Figure \ref{fig:vorticity_terms} $(b)$ shows that vorticity formation for the entire $Ra$ range is mainly governed by buoyancy and balanced by diffusion. Assume the hot plumes carry warm fluid to the upper plate where it meets a cold sidewall, generating strong lateral gradients in the upper right corner and consequently vorticity, according to eq. \eqref{eq:vort}. 

In the adiabatic case, on the other hand, the sidewall is warmer close to the corner, which leads to less vorticity generation by lateral temperature gradients and therefore smaller corner rolls. In the low $Ra$ regime, the corner rolls of the adiabatic sidewall are also governed by buoyancy, with a growth of the corner rolls of $\delta_{CR}\sim Ra^{0.21}$ (figure \ref{fig:cr_size} $a$). This can be understood by dimensional arguments. Assume convection can be neglected in eq. \eqref{eq:vort}, which is justified from the results in figure \ref{fig:vorticity_terms} $(a)$. Thus we obtain $\sqrt{Pr/Ra} \nab^2 \omega = \partial_x \theta$, or, in terms of a characteristic temperature $\theta_{CR}$ and a characteristic vorticity $\Omega_{CR}$, we have $\nu \frac{\Omega_{CR}}{\delta_{CR}^2} \sim \frac{\theta_{CR}}{\delta_{CR}}$, and thus
\begin{equation}
    \delta_{CR} \sim \sqrt{\frac{Pr}{Ra}} \frac{\Omega_{CR}}{\theta_{CR}}.
    \label{eq:cr_size}
\end{equation}
The evaluation (not shown here) of the characteristic vorticity in the corner roll regions by means of their root mean square value unveiled $\Omega \sim Ra^{0.7}$. Assuming further that the temperature $\theta_{CR}$ is approximately constant over $Ra$, we obtain $\delta_{CR}\sim Ra^{0.20}$, which agrees remarkably well with $\delta_{CR}\sim Ra^{0.21}$. Figure \ref{fig:cr_size} $(a)$ discloses a transition at $Ra \approx 3 \times 10^5$ , above which the corner roll growth accelerates exhibiting a scaling of $\delta_{CR}\sim Ra^{0.49}$. Figure \ref{fig:vorticity_terms} $(a)$ indicates that convective processes begin to affect vorticity generation. Figure \ref{fig:vorticity_fields} $(d)$ reveals a region with strong convective vorticity current with the same sign as the buoyancy forces, which enhances the vorticity generation in this region (figure \ref{fig:vorticity_fields} $c$). We interpret that above a certain $Ra$ the primary roll of the SRS begins to feed the corner rolls until they become strong enough, eventually leading to the collapse of the SRS itself. We would like to note that the current analysis describes steady-states up to $Ra\leq 10^6$. An opposite trend was observed for larger $Ra$ by \cite{Zhou2018}, who found a slow shrinkage of the corner rolls that scales approximately with $\sim Ra^{-0.085}$. It would be interesting to consolidate these results in future studies.

\subsection{Double-roll ($\mathcal{S}_A^2$, $\mathcal{S}_L^2$)}
\label{sec:roll2}

\begin{figure}
\unitlength1truecm
\begin{picture}(12, 5.0)
\put(2.0, 0.0){\includegraphics[height=4.6cm]{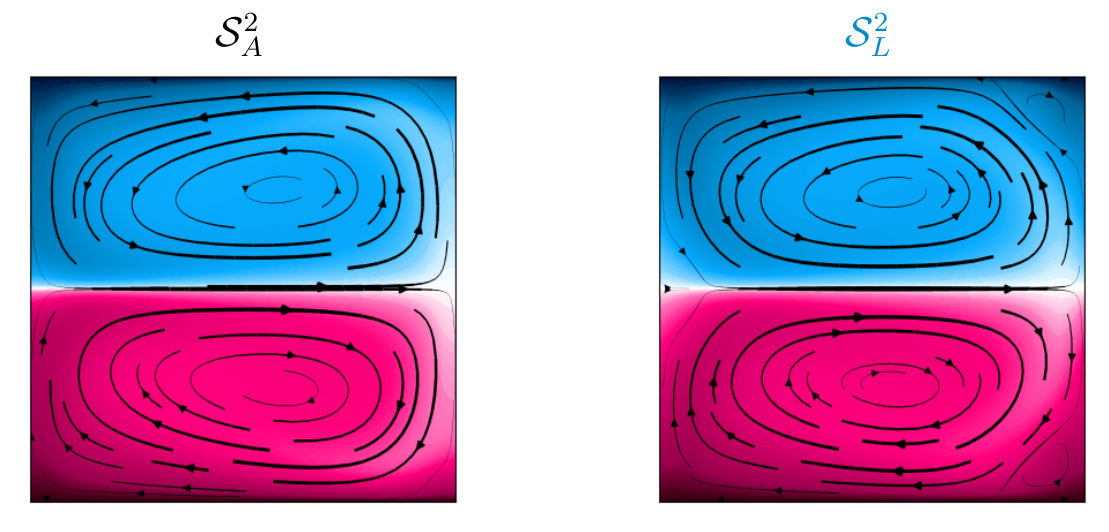}}
\put(1.5, 4.2){$(a)$}
\put(7.0, 4.2){$(b)$}
\put(1.8,1){\rotatebox{90}{Adiabatic SW}}
\put(7.25,1.3){\rotatebox{90}{Linear SW}}
\end{picture}
\caption{
Double-roll state (DRS) for $(a)$ adiabatic and $(b)$ linear. Contours (streamlines) represent the temperature (velocity) field.
}
\label{fig:steady_fields_roll2}
\end{figure}
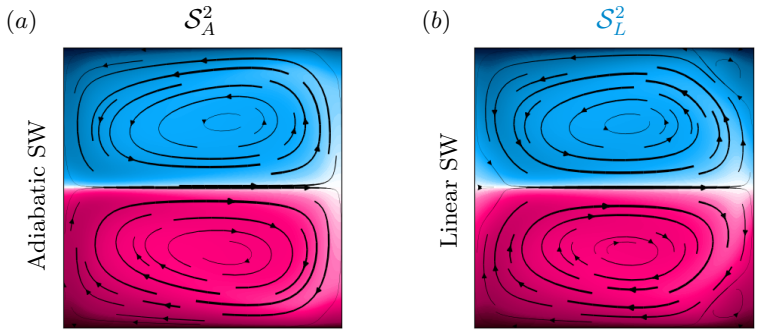

Having discussed the properties of the SRS state, we proceed to the double-roll state (DRS) as shown in figure \ref{fig:steady_fields_roll2}. It consists of two vertically stacked hot and cold circulation cells rotating in opposite directions with an almost discrete temperature jump in the mid plane. The DRS was not identified as an equilibrium for the constant sidewall BCs, so we will discuss it exclusively for the adiabatic and linear sidewall setup. The DRS can coexist with the SRS, but is generally found at larger $Ra$. Here we have tracked it in the range $10^5 \leq Ra < 7\times 10^6$ for adiabatic and $10^5 \leq Ra < 4\times 10^6$ for linear sidewall BCs. This range is consistent with \cite{Goldhirsch1989} who described a roll-upon-roll state in 2D RBC for $Pr=0.71$ at $Ra\approx10^5$, but interestingly it was not found for $Pr=6.8$.

\begin{figure}
\unitlength1truecm
\begin{picture}(12,5.4)
\put(0.8, 0.0){\includegraphics[height=4.6cm]{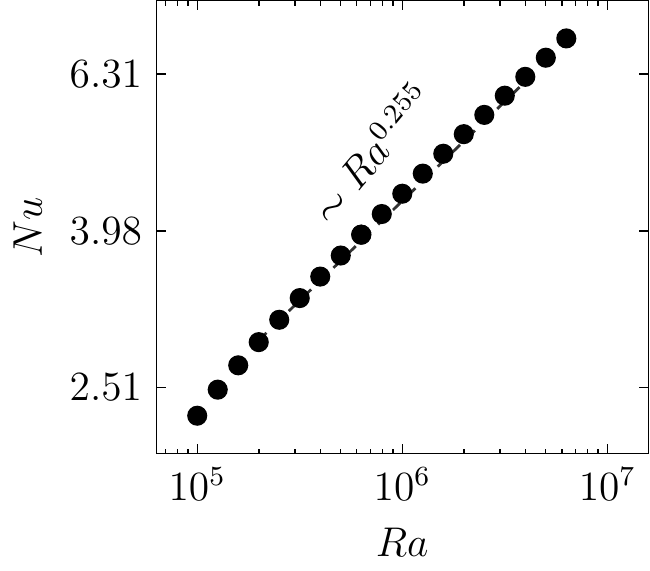}}
\put(7.02, 0.0){\includegraphics[height=4.6cm]{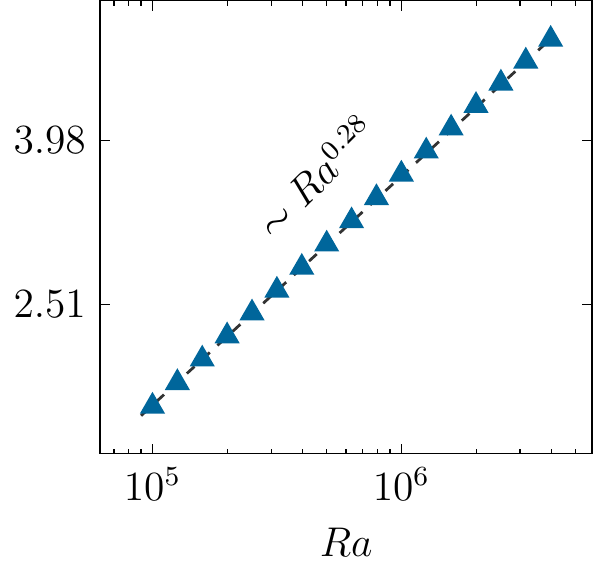}}
\put(1.3, 4.8){$(a)$}
\put(7.0, 4.8){$(b)$}
\put(3.0, 4.8){Adiabatic SW}
\put(9.2, 4.8){Linear SW}
\end{picture}
\caption{
Nusselt number $Nu$ for double-roll states $\mathcal{S}_A^2$ and $\mathcal{S}_L^2$. $(a)$ adiabatic and $(b)$ linear sidewall temperature boundary conditions.
}
\label{fig:roll2_nu}
\end{figure}
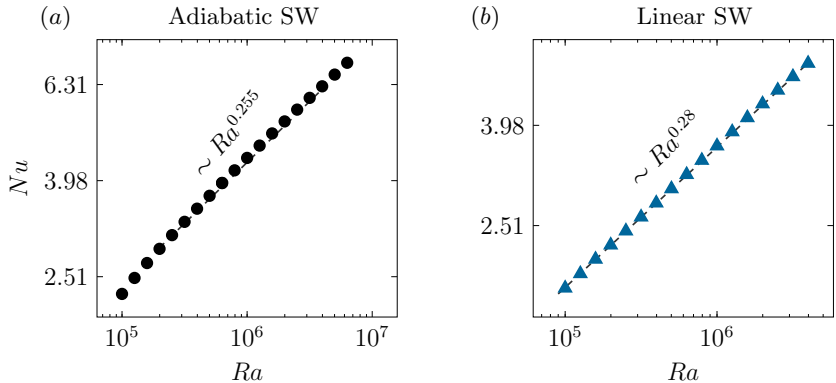

\begin{figure}
\unitlength1truecm
\begin{picture}(12, 5.4)
\put(0.9, 0.0){\includegraphics[height=4.6cm]{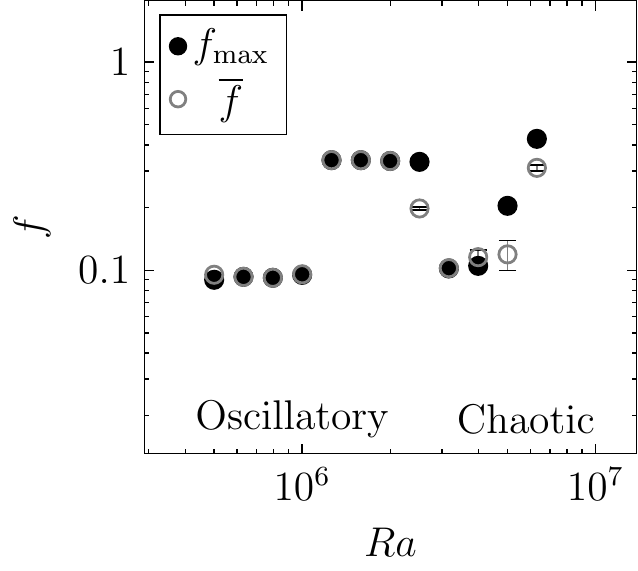}}
\put(7.0, 0.0){\includegraphics[height=4.6cm]{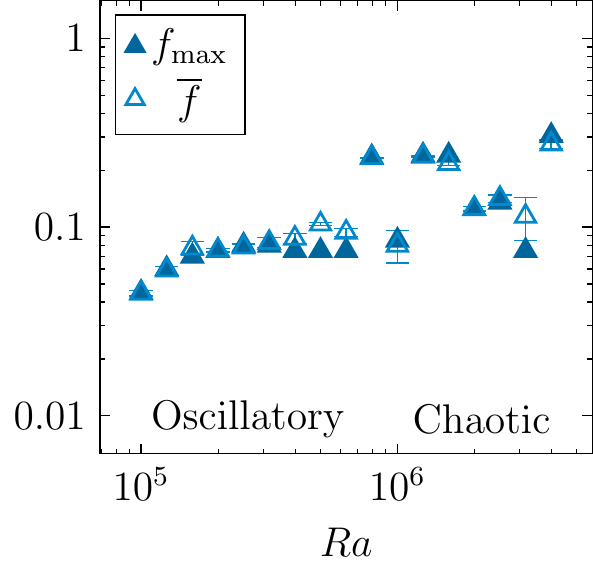}}
\put(1.3, 4.8){$(a)$}
\put(7.0, 4.8){$(b)$}
\put(3.0, 4.8){Adiabatic SW}
\put(9.2, 4.8){Linear SW}
\end{picture}
\caption{
Maximum peak frequency $f_{\text{max}}$ and average frequency $\overline{f}$ determined from $Nu(t)$ for double-roll states $\mathcal{S}_A^2$ and $\mathcal{S}_L^2$ for $(a)$ adiabatic and $(b)$ linear sidewall temperature boundary conditions.
}
\label{fig:roll2_f}
\end{figure}
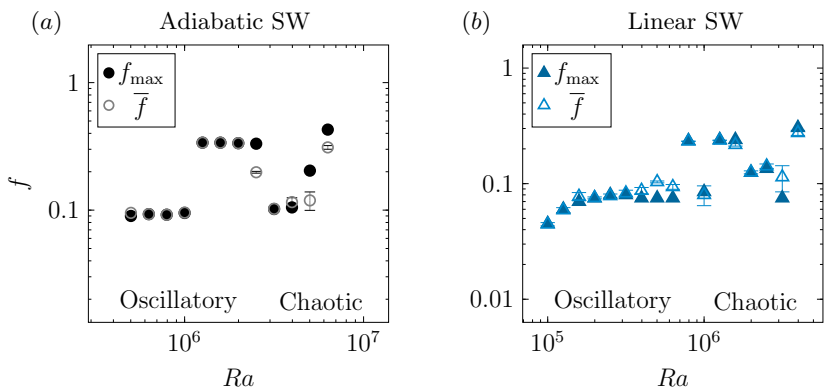

From figure \ref{fig:roll2_nu} we see that $Nu$ scales close to $Nu\sim Ra^{1/4}$, which corresponds to laminar scaling for RBC flows governed by boundary layer dissipation. Compared to the single-roll state, it is less effective in transporting heat from wall to wall, as evidenced by an overall smaller $Nu$. This is actually to be anticipated, since one roll of the DRS can be conceptually viewed as a half-height, half-temperature gradient RBC system, implying a $16$ times smaller effective $Ra$. However, this factor most likely overestimates the difference, since the mid plane velocity is much closer to a free-slip flow than a no-slip flow and the aspect ratio is two rather than one. In reality, a DRS has about the same $Nu$ as a SRS with a $6$ times smaller $Ra$. 

The DRS is found to be time-independent (stable) only for the adiabatic sidewall BCs for $Ra\leq 4 \times 10^5$. For other $Ra$ it is either periodically oscillating or chaotic. In figure \ref{fig:roll2_f} we show characteristic frequencies of the DRS obtained by initializing DNS simulation with the steady-state solutions and evaluating the frequency spectra of $Nu(t)$. The frequency is presented in free-fall time units. The DRS oscillates with a frequency of about $0.1$ for $Ra\leq 10^6$ for both the adiabatic and linear setups, i.e., about one cycle every $10$ time units. This cycle corresponds to about half the circulation time of a cell, i.e., the characteristic velocity of the circulation is about $0.09 \sim 0.11$ and its size is $\approx 2 L$. Thus, the DRS oscillation frequency seems to be initially tied to the circulation time. When $Ra$ exceeds $10^6$, we see the emergence of a more chaotic behavior. Despite increasing turbulence, the DRS state persists and does not show transition to a SRS state for $Ra<10^7$. In section \ref{sec:modal} we will see that for larger $Ra$ the DRS state is eventually replaced by a single roll LSC again.  

\begin{figure}
\unitlength1truecm
\begin{picture}(12, 5.6)
\put(0.1, 0.0){\includegraphics[height=4.9cm]{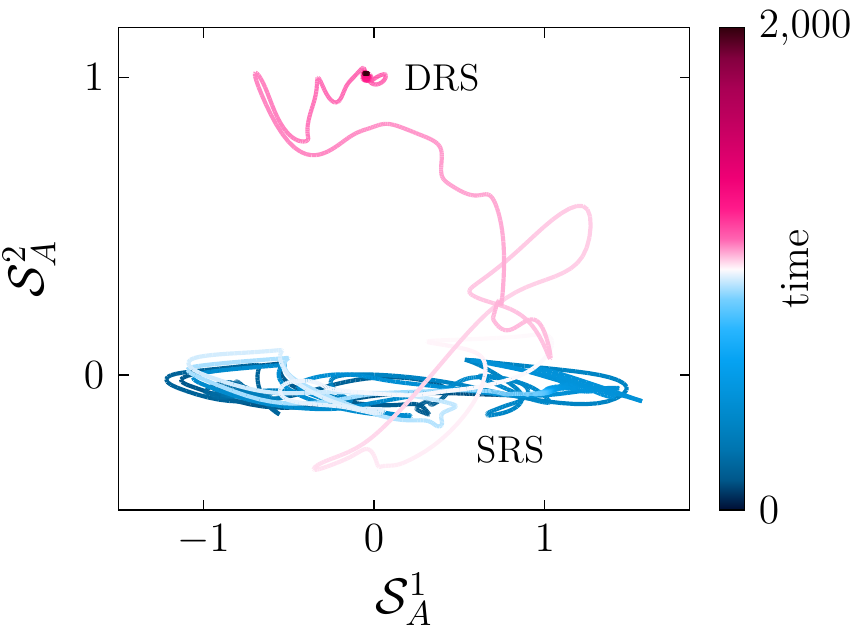}}
\put(7.0, 0.0){\includegraphics[height=4.9cm]{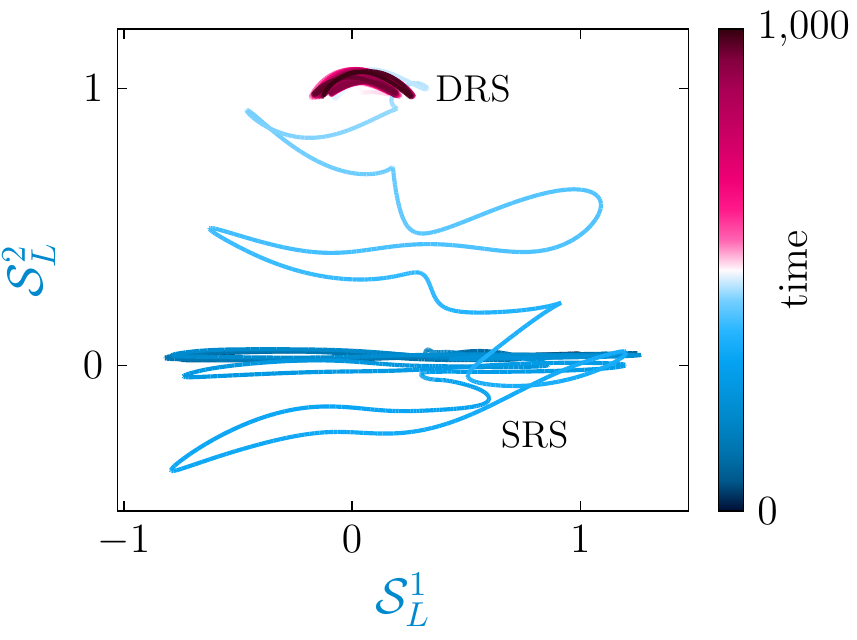}}
\put(0.0, 5.0){$(a)$}
\put(7.0, 5.0){$(b)$}
\put(2.2, 4.9){Adiabatic SW}
\put(9.4, 4.9){Linear SW}
\end{picture}
\caption{Phase space trajectories from a single-roll ($\mathcal{S}_A^1$/$\mathcal{S}_L^1$) to a double-roll state ($\mathcal{S}_A^2$/$\mathcal{S}_L^2$) for $(a)$ adiabatic sidewall at $Ra=2\times10^6$ and $(b)$ linear sidewall BCs at $Ra=1.5\times10^5$.}
\label{fig:roll1_to_roll2}
\end{figure}
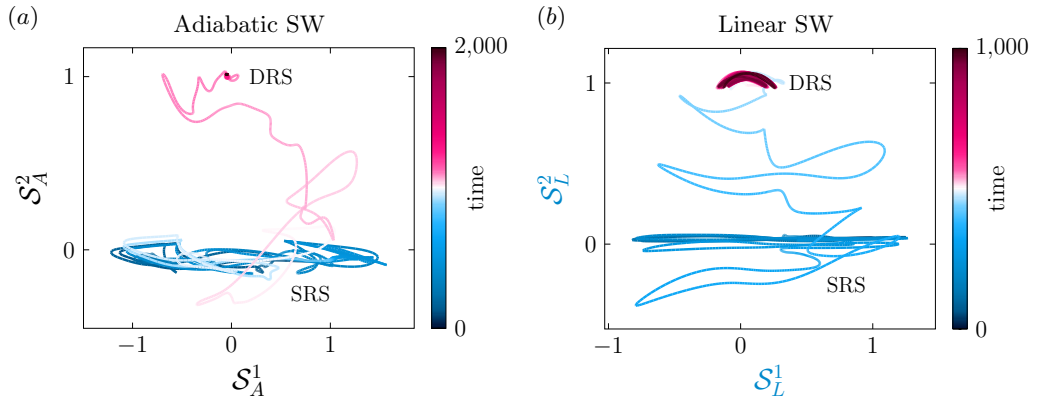

The DRS state is not merely an equilibrium solution, but more fundamentally there is a regime in $Ra$ where the DRS is the preferred flow state to which all initial states tested in this work tend towards. Starting from random perturbations, one usually first finds a SRS, which soon goes through a series of flow reversals and restabilizations until it evolves to the DRS state. This process is depicted in an SRS-DRS phase space picture in figure $\ref{fig:roll1_to_roll2}$. The horizontal axis represents the SRS, and the vertical axis represents the DRS. This process is qualitatively the same for adiabatic and linear sidewall boundary conditions. We do not address the flow reversal process, as it is described in more detail in \cite{Xi2007, Sugiyama2010, Castillo2016, Zhao2019}, but note that the intermediate flow fields bear striking resemblance to the proper orthogonal decomposition modes presented in \cite{Podvin2015, Podvin2017}. We want to stress that the transition time is surprisingly long. It can take up to several thousand free-fall time units for the flow to settle in the DRS state, so it can be missed if the observation window is too small.

\section{Direct numerical simulations}

In addition to the steady-state analysis, we performed a series of DNS of RBC for 2D in a square and 3D in a cylinder with $\Gamma=1$ and $Pr=1$, covering $Ra$ from the onset of convection to $4.64 \times 10^{10}$ and $10^9$, respectively. The highest $Ra$ in 2D was simulated on a $1024^2$ grid with at least $15$ grid points in the thermal boundary layer and performed for several thousand free-fall time units, ensuring adequate spatial resolution and temporal convergence. The largest simulation for the cylindrical setup was performed on a $N_r \times N_\varphi \times N_z = 128 \times 256 \times 320$ grid, with about $10$ points inside the thermal and viscous boundary layers and the averaging statistics were collected for at least $600$ free-fall time units.

\subsection{Vertical temperature profiles}

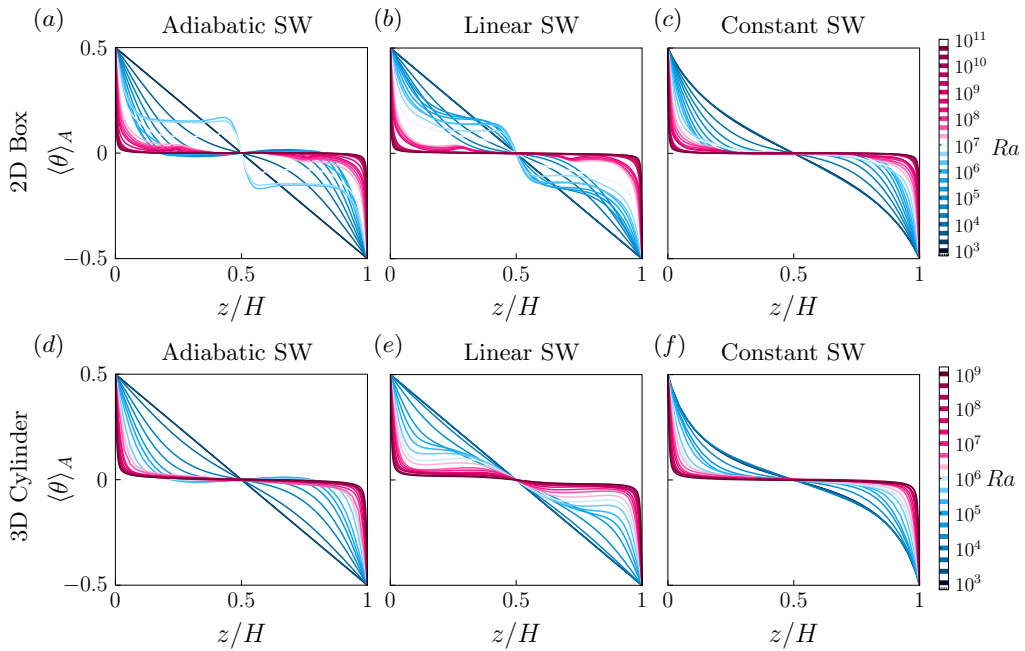
\begin{figure}
\unitlength1truecm
\begin{picture}(12, 10)
\put(0.4, 4.3){\includegraphics[height=3.9cm]{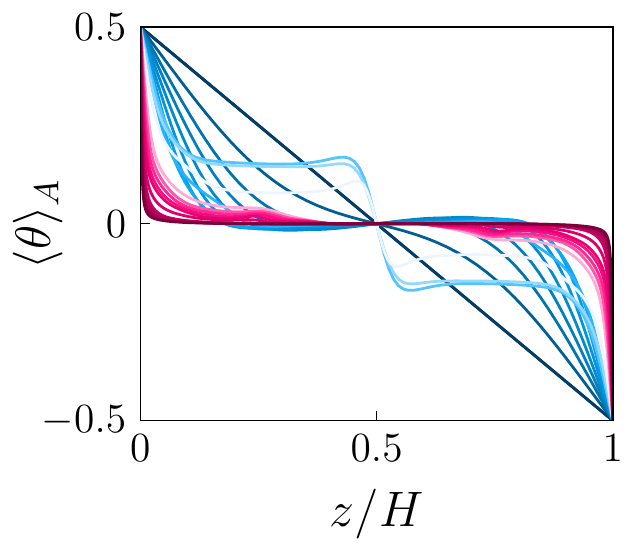}}
\put(4.83, 4.3){\includegraphics[height=3.8cm]{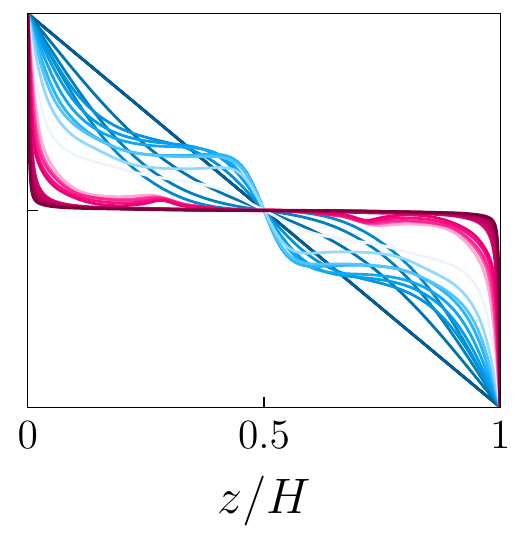}}
\put(8.5, 4.3){\includegraphics[height=3.8cm]{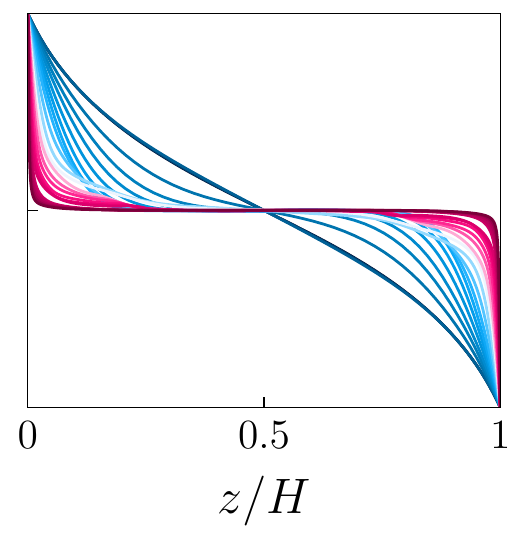}}
\put(12.2, 5.1){\includegraphics[height=3.2cm]{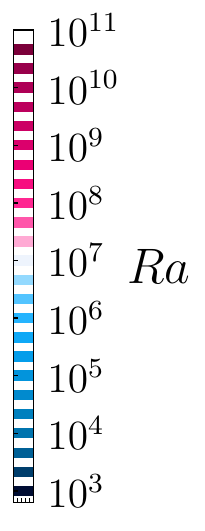}}
\put(0.3, 8.3){$(a)$}
\put(4.8, 8.3){$(b)$}
\put(8.5, 8.3){$(c)$}
\put(0.0,6.1){\rotatebox{90}{2D Box}}
\put(2.0, 8.2){Adiabatic SW}
\put(6.0, 8.2){Linear SW}
\put(9.4, 8.2){Constant SW}
\put(0.4, 0){\includegraphics[height=3.9cm]{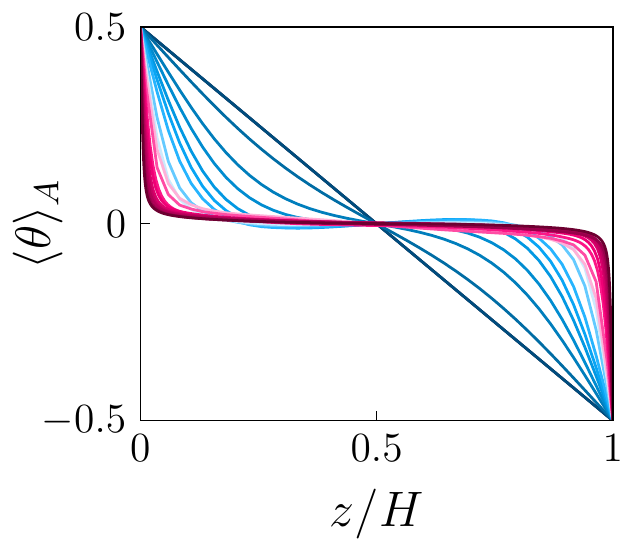}}
\put(4.83, 0){\includegraphics[height=3.8cm]{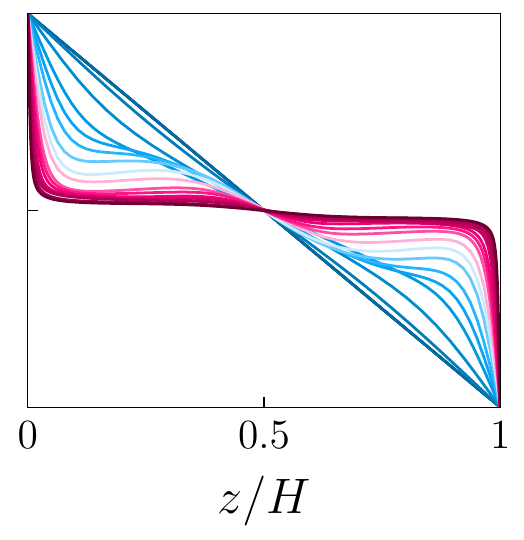}}
\put(8.5, 0){\includegraphics[height=3.8cm]{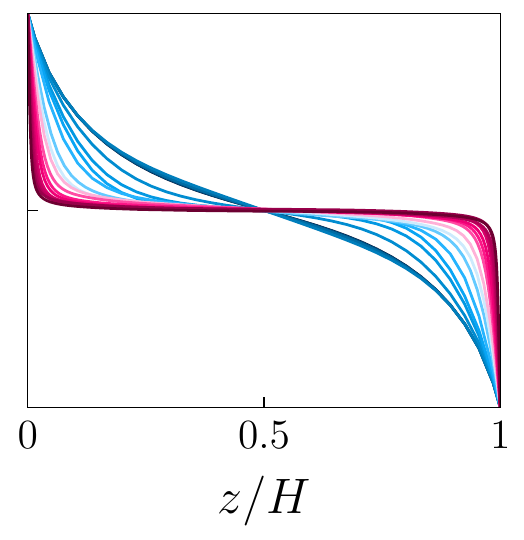}}
\put(12.2, 0.7){\includegraphics[height=3.2cm]{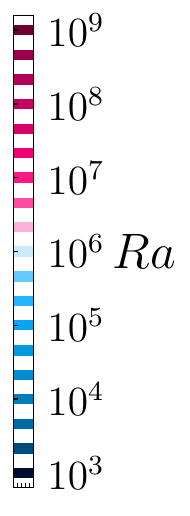}}
\put(0.3, 4.0){$(d)$}
\put(4.8, 4.0){$(e)$}
\put(8.5, 4.0){$(f)$}
\put(0.0,1.5){\rotatebox{90}{3D Cylinder}}
\put(2.0, 3.9){Adiabatic SW}
\put(6.0, 3.9){Linear SW}
\put(9.4, 3.9){Constant SW}
\end{picture}
\caption{
Mean temperature profile for cases with $(a,d)$ adiabatic, $(b,e)$ linear and $(c, f)$ constant sidewall boundary conditions for ($a$-$c$) 2D box and ($d$-$f$) cylinder. 
}
\label{fig:profiles_t}
\end{figure}

Figure \ref{fig:profiles_t} shows the horizontally averaged temperature profiles $\langle \theta \rangle_A$ for all conducted simulations. We first remark the similarity between 2D and 3D. For example, both show the feature of a weakly stabilizing positive temperature gradient in the mid plane for small $Ra$ and adiabatic boundary conditions (figures \ref{fig:profiles_t} a,d). This phenomenon is often found in the interior of the bulk \citep{Tilgner1993, Brown2007a, Wan2019} and is caused by the thermal signature of the LSC. As the thermal plume of the LSC climbs up along the sidewall, it penetrates deeper into the bulk, thus hot (cold) plumes carry their signature into the top (bottom) part of the cell, which can result in a slightly positive temperature gradient in the center of the bulk.

Another important detail is the apparent non-monotonicity of the profiles in the intermediate $Ra$ range, which is most pronounced for the linear sidewall BCs (figure \ref{fig:profiles_t} b,e) and also occurs for the 2D adiabatic BCs. The temperature profiles initially drop sharply and then level of at about a quarter of the cell height before dropping sharply again in the cell center. This behaviour was also observed in \cite{Stevens2014}. These profiles are reminiscent of the DRS state (see section \ref{sec:roll2}) and indeed caused by transitions in the flow structures, which we analyse in section \ref{sec:modal} in more detail. Finally, all simulations for larger $Ra$ show the classical RBC profile with steep temperature gradients at the bottom and top plates and a well-mixed homogeneous bulk.

\subsection{Vertical sidewall heat flux profiles}
\label{sec:dtdr}

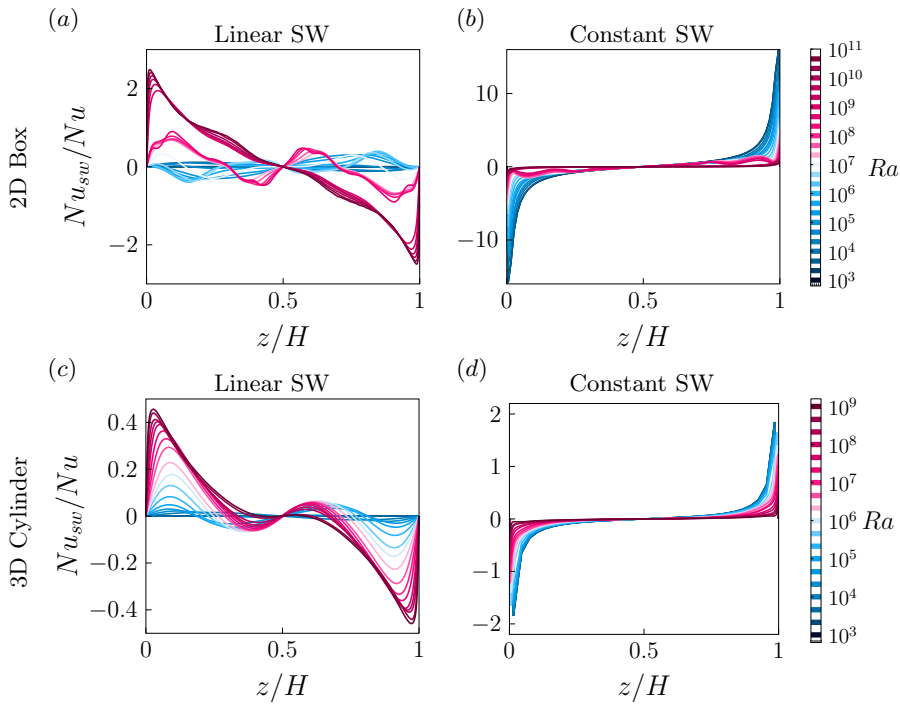
\begin{figure}
\unitlength1truecm
\begin{picture}(12, 10)
\put(1.6, 4.6){\includegraphics[height=4.1cm]{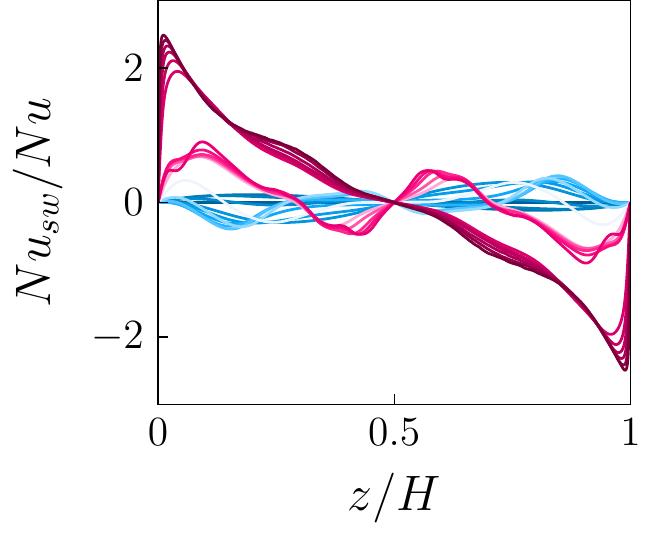}}
\put(6.8, 4.6){\includegraphics[height=4.1cm]{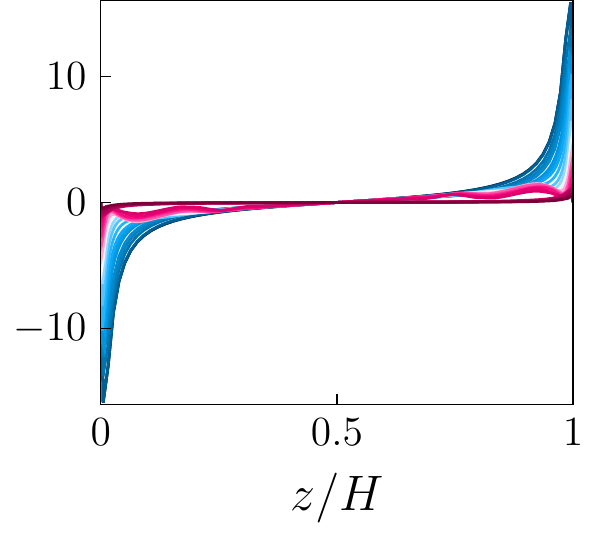}}
\put(11.5, 5.4){\includegraphics[height=3.5cm]{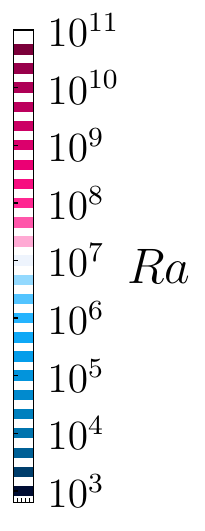}}
\put(1.5, 9.0){$(a)$}
\put(6.9, 9.0){$(b)$}
\put(1.0,6.6){\rotatebox{90}{2D Box}}
\put(3.7, 8.8){Linear SW}
\put(8.4, 8.8){Constant SW}
\put(1.5, 0.0){\includegraphics[height=4.1cm]{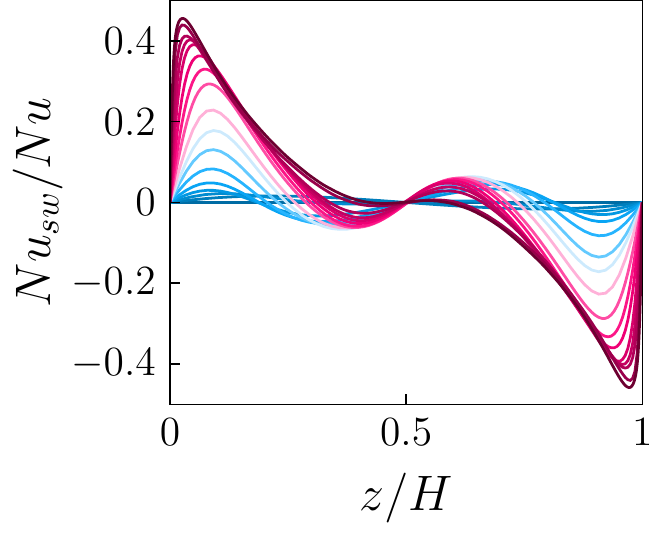}}
\put(7.0, 0.0){\includegraphics[height=4.1cm]{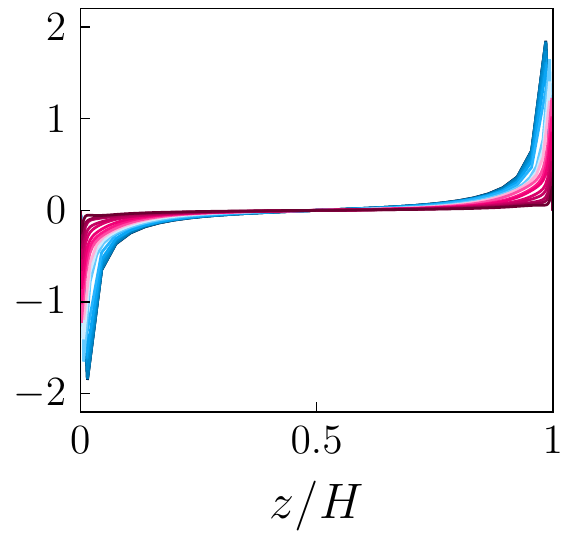}}
\put(11.5, 0.7){\includegraphics[height=3.5cm]{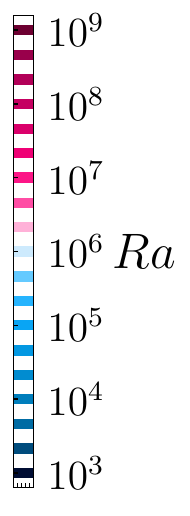}}
\put(1.5, 4.4){$(c)$}
\put(6.9, 4.4){$(d)$}
\put(1.0,1.6){\rotatebox{90}{3D Cylinder}}
\put(3.7, 4.2){Linear SW}
\put(8.4, 4.2){Constant SW}
\end{picture}
\caption{
Comparison of the lateral sidewall heat flux $Nu_{sw}$ for cases ($a, c$) linear and ($b, d$) constant sidewall boundary conditions in ($a,b$) 2D box and ($c,d$) cylinder. 
}
\label{fig:profiles_dtdr}
\end{figure}

Next we analyse the horizontal heat flux through the vertical sidewall $Nu_{sw}$ which is more elaborately defined in the appendix \ref{app:heat_flux}. This is shown in figure \ref{fig:profiles_dtdr} for the linear and constant BCs, while the sidewall heat flux of the adiabatic BC is obviously zero. The linear and constant BCs show two opposite trends. The constant setup has the largest temperature gradients for small $Ra$ and almost vanishing gradients for large $Ra$. This can be understood from the temperature profiles in figure \ref{fig:profiles_t} $(c, f)$. As $Ra$ increases, the bulk is more efficiently mixed and the temperature distribution becomes nearly constant, hence the temperature in the cell becomes more similar to the sidewall temperature imposed by the BCs. On the other hand, the linear sidewall BC corresponds exactly to the temperature profile before the onset of convection and from then on its contrast increases more and more, which is reflected in the relatively strong vertical temperature gradients for large $Ra$. However, all profiles are symmetrical around the center and consequently, although heat flows in and out locally, there is no net heat flux through the vertical sidewalls. This is supported by the fact that in our simulations the temperature gradients at the top and bottom plates were nearly equal, linked by the heat flux balance 
\begin{equation}
   Nu_c - Nu_h + \zeta \langle Nu_{sw} \rangle_z = 0 
\end{equation}
with $\zeta = \frac{1}{\Gamma}$ for the 2D box and  $\zeta = \frac{4}{\Gamma}$ for the cylindrical setup (see appendix \ref{app:heat_flux}). Lastly, we detect at least two transitions in $Nu_{sw}$ for the linear sidewall BCs (figure \ref{fig:profiles_dtdr} $a, c$). These are consistent with the transitions in the temperature profiles discussed in the previous section and are elucidated in more detail in the following.

\subsection{Mode analysis}
\label{sec:modal}

It is generally difficult to compare the dynamics of flows in different, possibly even turbulent, states without restricting the underlying state space. Therefore, in this section we analyze the DNS results by projecting each snapshot onto four distinct modes and evaluate time averages and standard deviations.

Starting with the 2D simulations, a common choice for the mode are the first four Fourier modes, see e.g. \cite{Petschel2011} and \citep{ Wagner2013}, i.e.
\begin{align}
    u_x^{m,k} &= -\sin(\pi m x/L) \cos(\pi k z/H), \nonumber\\
    u_z^{m,k} &= \cos(\pi m  x/L) \sin(\pi k z/H).
\label{eq:mode_box}
\end{align}
For the cylinder, the choice of modes is less obvious. In this work, we follow \cite{Shishkina2021} and use a combination of Fourier modes in $z$ and $\varphi$ direction and Bessel functions of the first kind $J_n$ of order $n$ in $r$ for the radial velocity component $u_r$ and the vertical velocity component $u_z$. The first two (non-axisymmetric) modes are
\begin{align}
    u_r^{1,k} &= J_0 (\alpha_{0} r/R) \cos(\pi k z/H) e^{i \varphi},\nonumber\\
    u_z^{1,k} &= J_1 (\alpha_{1} r/R) \sin(\pi k z/H) e^{i \varphi},
\label{eq:mode_cyl2}
\end{align}
and the axisymmetric modes are
\begin{align}
    u_r^{2,k} &= J_1 (\alpha_{1} r/R) \cos(\pi k z/H) ,\nonumber\\
    u_z^{2,k} &= -J_0 (\alpha_{0} r/R) \sin(\pi k z/H),
\label{eq:mode_cyl1}
\end{align}
where $\alpha_{n}$ is the first positive root of the Bessel function $J_n$ for Dirichlet boundary conditions on the sidewall ($u_r$) and the $k$-th positive root of the derivative of the Bessel function $J_n^\prime$ for Neumann boundary conditions $(u_z)$. The non-axisymmetric modes are complex-valued to account for different possible azimuthal orientations. Ultimately, however, we are only interested in the energy content and not the orientation of the modes, so we evaluate their magnitude. We note further, that a vertical slice through the cylindrical modes is very similar to the first four 2D Fourier modes, albeit with a slightly different dependence in the radial direction. For this reason, we use the same notation for the cylindrical modes as for the Fourier modes in 2D. More precisely, we have $F_1\equiv(u_r^{1,1},u_z^{1,1})$, $F_2^=\equiv(u_r^{1,2},u_z^{1,2})$, $F_2^\parallel \equiv(u_r^{2,1},u_z^{2,1})$ and $F_4 \equiv(u_r^{2,2},u_z^{2,2})$. Having defined the modes, we project the velocity field $\uu$ of several snapshots onto a mode $\uu^m$ and evaluate the energy content $\mathcal{P}$ of each mode according to
\begin{align}
    \mathcal{P} \equiv \frac{\int_V \uu \uu^m dV}{\int_V \uu^m \uu^m dV},
\label{eq:mode_energy}
\end{align}
and analyse the time average and standard deviation of $\mathcal{P}$.

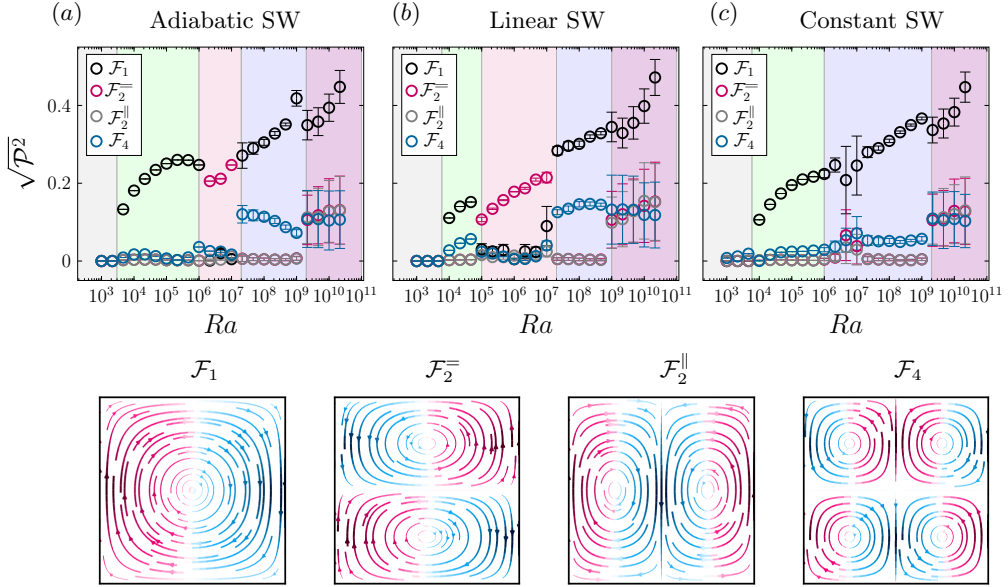
\begin{figure}
\unitlength1truecm
\begin{picture}(12, 8.0)
\put(1.05, 0.0){\includegraphics[width=2.6cm]{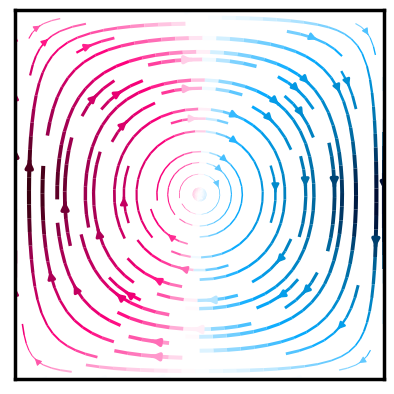}}
\put(4.15, 0.0){\includegraphics[width=2.6cm]{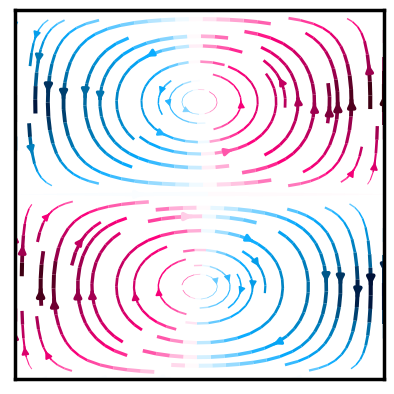}}
\put(7.25, 0.0){\includegraphics[width=2.6cm]{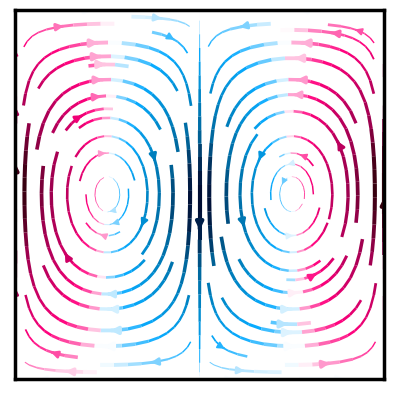}}
\put(10.35, 0.0){\includegraphics[width=2.6cm]{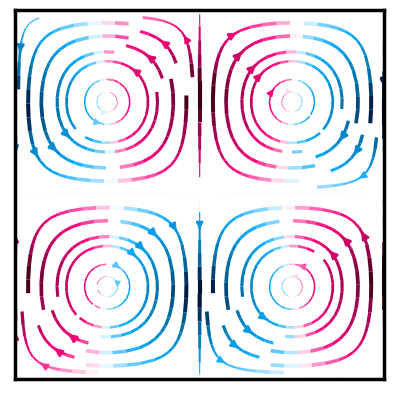}}
\put(2.35, 2.8){$\mathcal{F}_1$}
\put(5.45, 2.8){$\mathcal{F}_2^{=}$}
\put(8.55, 2.8){$\mathcal{F}_2^\parallel$}
\put(11.65, 2.8){$\mathcal{F}_4$}
\put(-0.1, 3.3){\includegraphics[height=3.9cm]{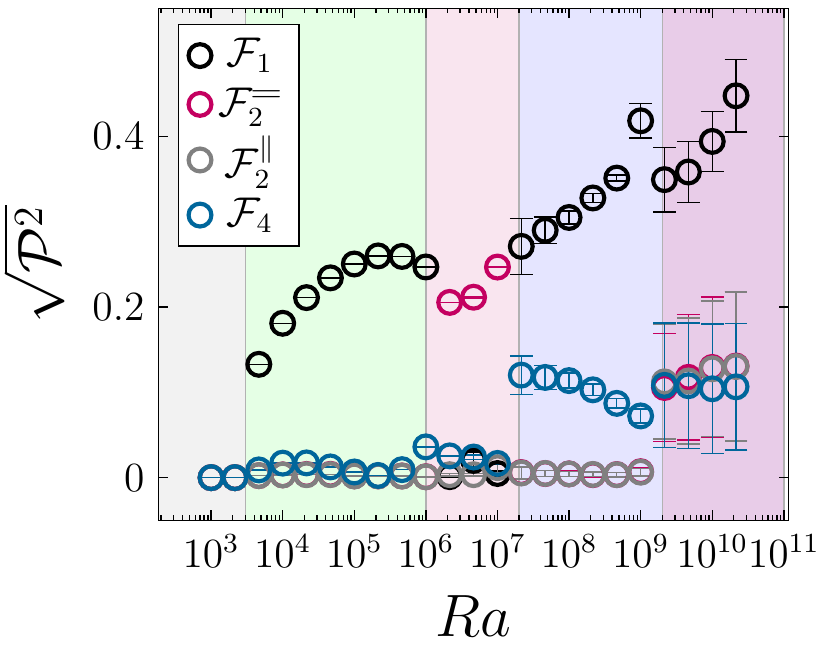}}
\put(5.0, 3.3){\includegraphics[height=3.9cm,trim={1.7cm 0 0 0},clip]{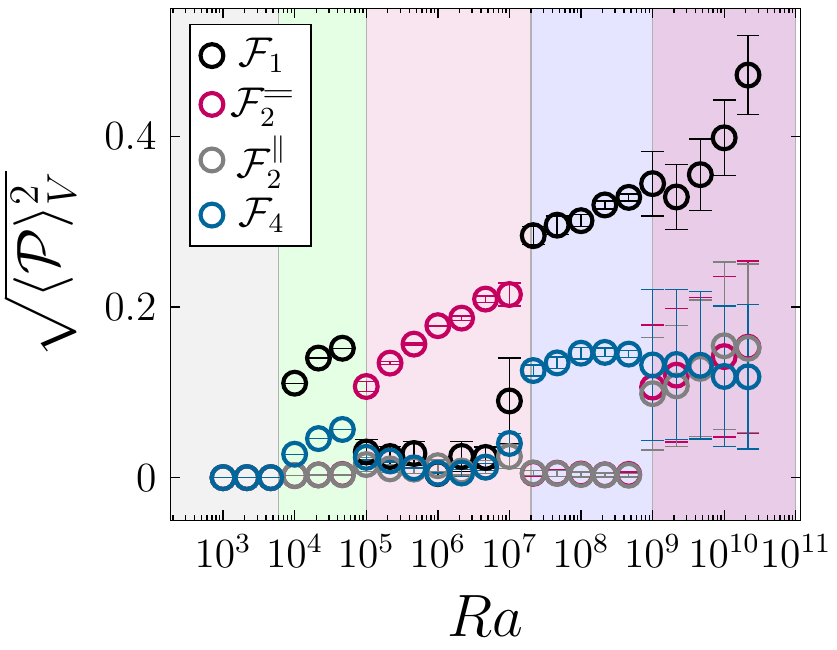}}
\put(9.1, 3.3){\includegraphics[height=3.9cm,trim={1.7cm 0 0 0},clip]{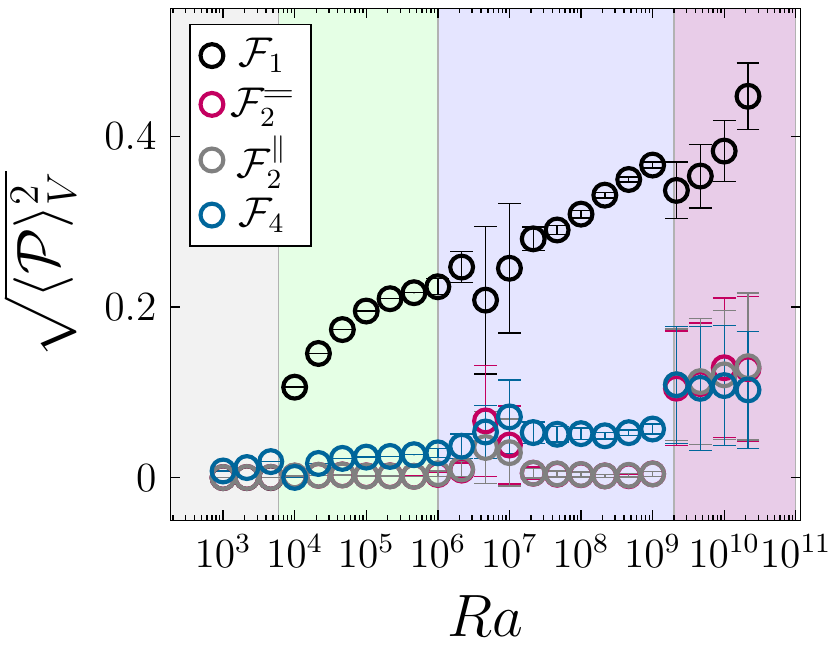}}
\put(0.5, 7.5){$(a)$}
\put(5.0, 7.5){$(b)$}
\put(9.2, 7.5){$(c)$}
\put(1.8, 7.4){Adiabatic SW}
\put(6.3, 7.4){Linear SW}
\put(10.4, 7.4){Constant SW}
\end{picture}
\caption{Energy and standard deviation of the projection of flow field snapshots onto the modes defined by eq. \eqref{eq:mode_box} for the 2D box and $(a)$ adiabatic, $(b)$ linear and $(c)$ constant sidewall temperature boundary condition for the 2D box. Below: Streamlines, coloured by vertical velocity, of the modes $\mathcal{F}_1$, $\mathcal{F}_2^{=}$, $\mathcal{F}_2^{\parallel}$ and $\mathcal{F}_4$.}
\label{fig:mode_box}
\end{figure}

The energy of the individual Fourier mode for the 2D box is shown in figure \ref{fig:mode_box}. Above the onset of convection, only the first Fourier mode (single-roll) contains a considerable amount of energy. Because of its similarity to the SRS, this mode will be referred to as the SRS-mode. Following the stable SRS, we find for adiabatic and linear sidewall BCs a flow regime that changes from the SRS to a roll-upon-roll second Fourier mode ($\mathcal{F}_2^{\parallel}$) state. This state embodies the DRS state, which we discussed in section \ref{sec:roll2}. The $F_2^=$ regime, or DRS regime, is found in the range $10^6 < Ra \leq 10^7$ for an adiabatic sidewall and $10^5 \leq Ra \leq 10^7$ for a linear sidewall BC. In contrast, the DRS regime is absent for a constant sidewall BC. As a reminder, this state could not be found as an equilibrium solution for the constant sidewall boundary condition either, which is in line with its absence in DNS. The next regime can be regarded as a weakly chaotic SRS regime, with the SRS mode again dominating but being transient and a substantial amount of energy is contained in the $F_4$ (4-roll) mode, indicative of dynamically active corner rolls. Finally, above $Ra \approx 10^9$ there exists another surprisingly sharp transition. This regime is different from the others as now all Fourier modes contain a significant amount of energy and exhibit strong fluctuations. An inspection of the flow fields revealed an abundance of small-scale plumes and strong turbulent dynamics. Most remarkably, in this regime all three sidewall BCs show a very similar mode signature, i.e., they become increasingly alike, or in other words, RBC becomes insensitive to sidewall BCs for large $Ra$.

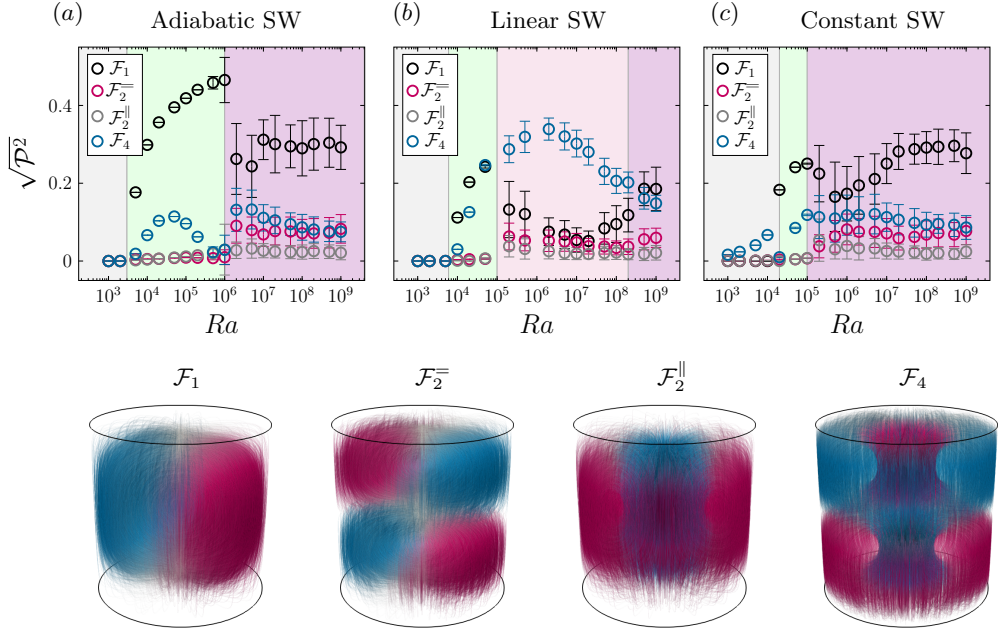
\begin{figure}
\unitlength1truecm
\begin{picture}(12, 8.5)
\put(-0.1, 3.8){\includegraphics[height=3.9cm]{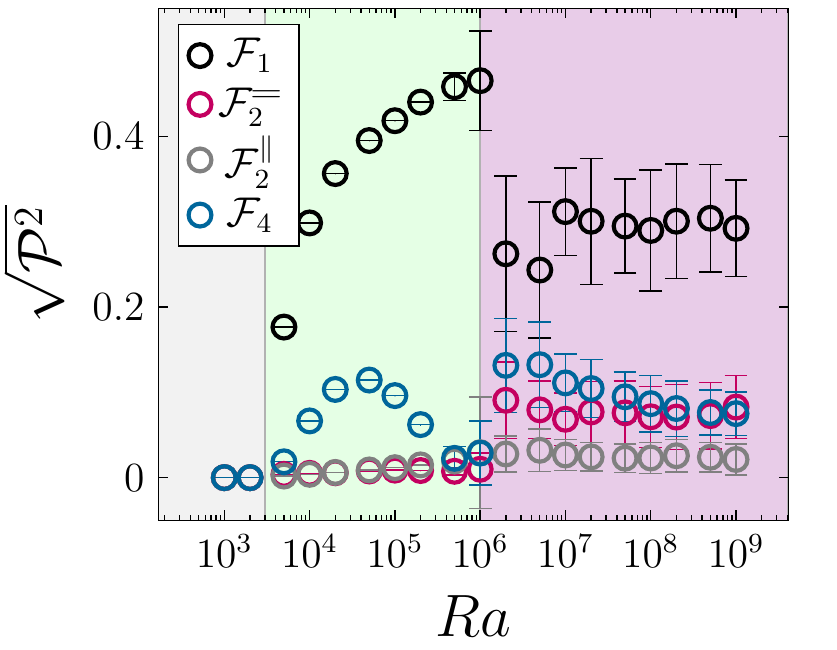}}
\put(5.0, 3.8){\includegraphics[height=3.9cm,trim={1.7cm 0 0 0},clip]{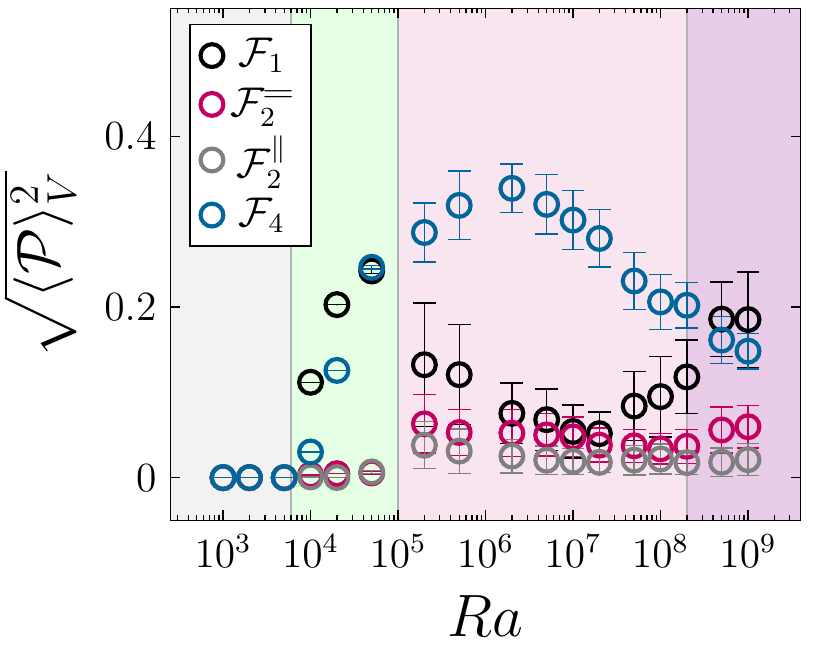}}
\put(9.1, 3.8){\includegraphics[height=3.9cm,trim={1.7cm 0 0 0},clip]{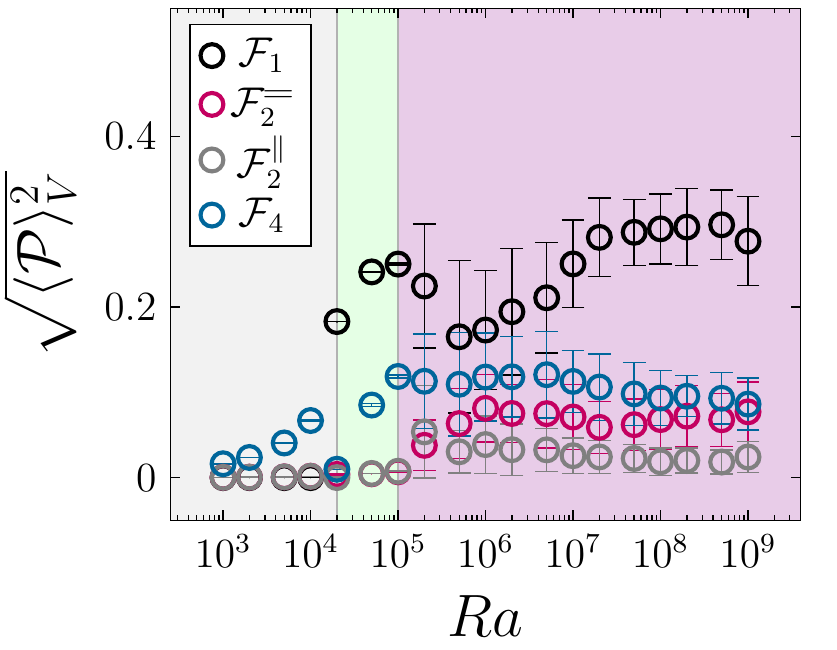}}
\put(0.5, 8.0){$(a)$}
\put(5.0, 8.0){$(b)$}
\put(9.2, 8.0){$(c)$}
\put(1.8, 7.9){Adiabatic SW}
\put(6.3, 7.9){Linear SW}
\put(10.4, 7.9){Constant SW}
\put(1.0, 0.0){\includegraphics[width=2.4cm,trim={5.5cm 2.6cm 5.5cm 3.8cm },clip]{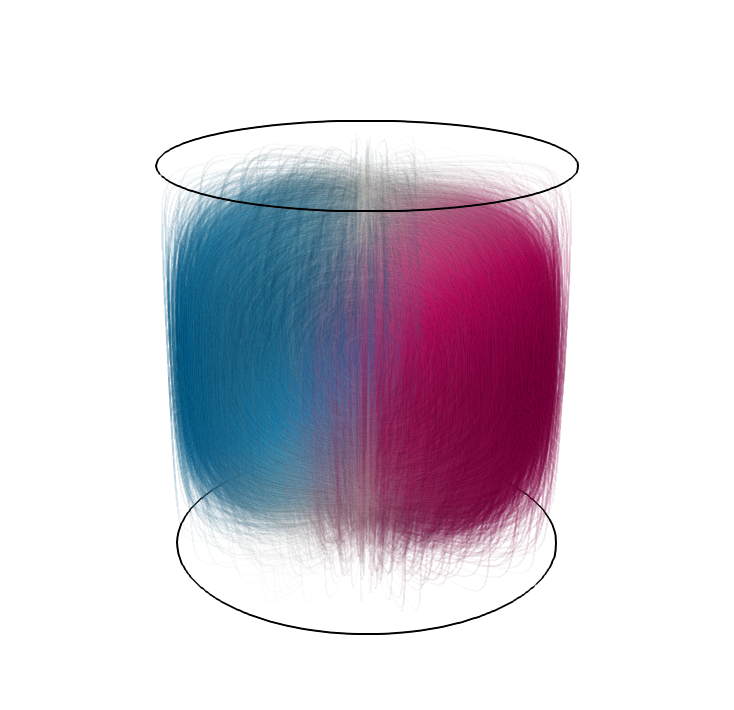}}
\put(4.2, 0.0){\includegraphics[width=2.4cm,trim={5.5cm 2.6cm 5.5cm 3.8cm },clip]{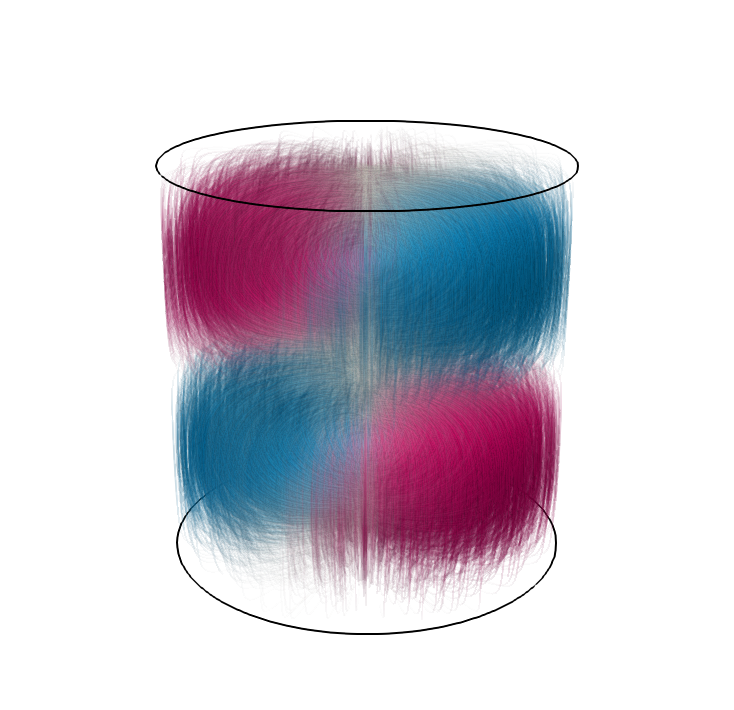}}
\put(7.4, 0.0){\includegraphics[width=2.4cm,trim={5.5cm 2.6cm 5.5cm 3.8cm },clip]{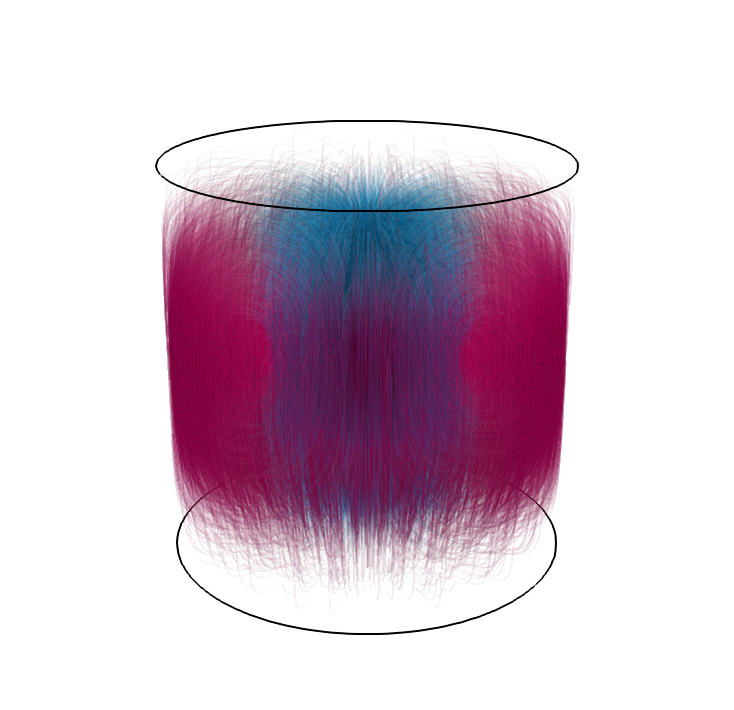}}
\put(10.6, 0.0){\includegraphics[width=2.4cm,trim={5.5cm 2.6cm 5.5cm 3.8cm},clip]{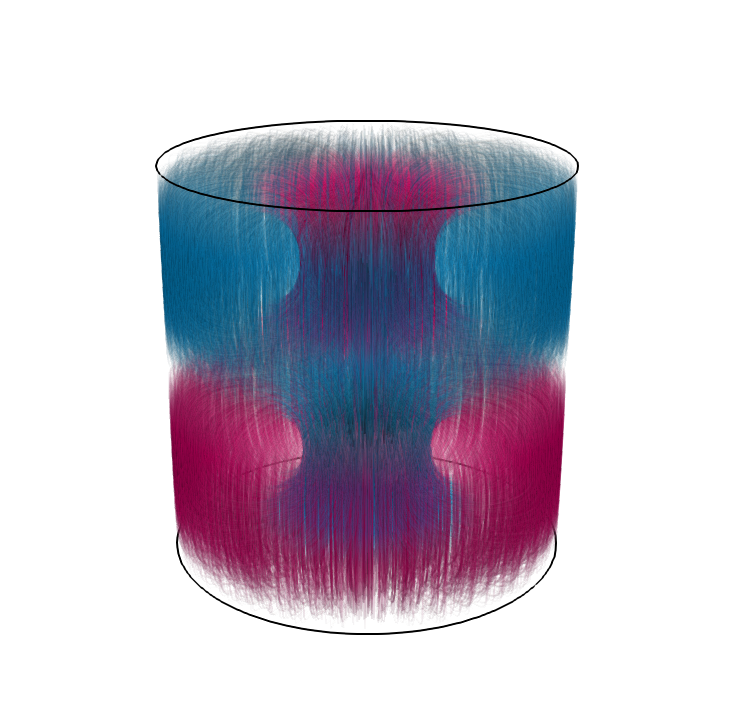}}
\put(2.1, 3.2){$\mathcal{F}_1$}
\put(5.3, 3.2){$\mathcal{F}_2^{=}$}
\put(8.5, 3.2){$\mathcal{F}_2^\parallel$}
\put(11.7, 3.2){$\mathcal{F}_4$}
\end{picture}
\caption{Energy and standard deviation of the projection of flow field snapshots onto the modes defined by eq. \eqref{eq:mode_cyl1} and \eqref{eq:mode_cyl2} for $(a)$ adiabatic, $(b)$ linear and $(c)$ constant sidewall temperature boundary condition for the cylinder. Below: Streamlines, coloured by vertical velocity, of the modes $\mathcal{F}_1$, $\mathcal{F}_2^{=}$, $\mathcal{F}_2^{\parallel}$ and $\mathcal{F}_4$.}
\label{fig:mode_cyl}
\end{figure}

Moving on to the mode analysis for the cylindrical setup, shown in figure \ref{fig:mode_cyl}, we see a very similar picture as for the 2D box with some noticeable differences. First, for the constant BC setup we note that the onset of convection is significantly later than in the 2D case, while the other two setups show a closer similarity with the 2D case. The cylindrical setup might be more sensitive to the BCs of the sidewalls in general, since the ratio of sidewall area to cell volume ratio is larger than in the 2D box and therefore the sidewall temperature likely has a larger impact on the interior. 

Another difference between the cylindrical and 2D box setup is, that the adiabatic setup does not show a transition to a regime with a vanishing SRS; rather, the SRS mode is the most dominant mode over all $Ra$. In contrast, the linear sidewall BC possess a striking similarity to the observations in 2D. Above $Ra\approx10^5$ it undertakes a transition from a SRS-dominated regime to a $F_4$-dominated regime. The $F_4$-mode is axissymmetric and has a double-donut, or double-toroidal shape. Similar flow states were found in a bifurcation analysis by \cite{Puigjaner2008} in a cubic domain with the same lateral boundary conditions. Here, its existence range extends over $10^5 \leq Ra \leq 10^8$. The double-donut state can be considered as the counterpart of the DRS state in 2D RBC, although we see that it outlasts its 2D analog by about a decade in $Ra$. At the highest $Ra$ available, the SRS again dominates for all BC configurations considered, although the amount of energy and the strength of the fluctuations are somewhat different for the different BCs. At this points, we can only conjecture from their trend and our findings in 2D that their deviation will decrease for even larger $Ra$ in the high-turbulence/high-$Ra$ regime.

We conclude that there exist at least five different flow regimes: conduction state, stable SRS, DRS (or double-donut state in the cylindrical setup), weakly chaotic SRS and highly turbulent state. We find the constant isothermal sidewall generally enhances the SRS dominance, while a linear isothermal sidewall BC suppresses the SRS in the mid $Ra$ regime and induces the DRS or double-donut state. Moreover, although we find strong differences in the flow dynamics in the small to medium $Ra$ range, but these differences eventually disappear and the system becomes increasingly insensitive to the type of sidewall BC at high $Ra$.

\subsection{Heat transport}
\label{sec:heat_transport}

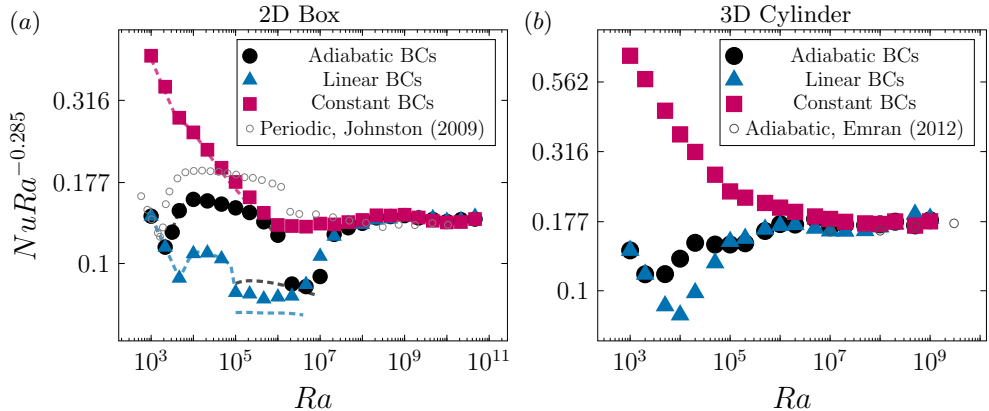
\begin{figure}
\unitlength1truecm
\begin{picture}(12, 5.8)
\put(0.0, 0.2){\includegraphics[height=5.3cm]{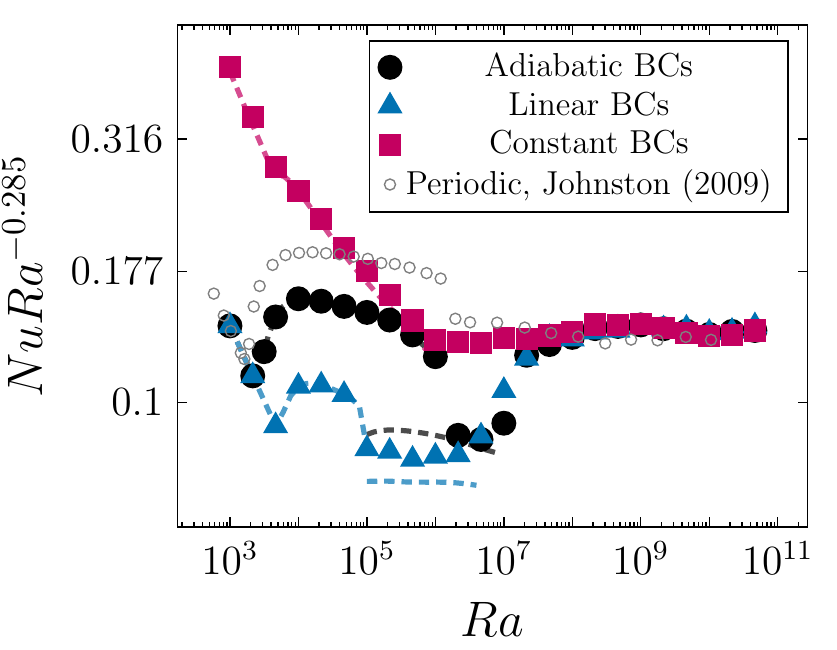}}
\put(6.9, 0.2){\includegraphics[height=5.3cm, trim=20 0 0 0, clip]{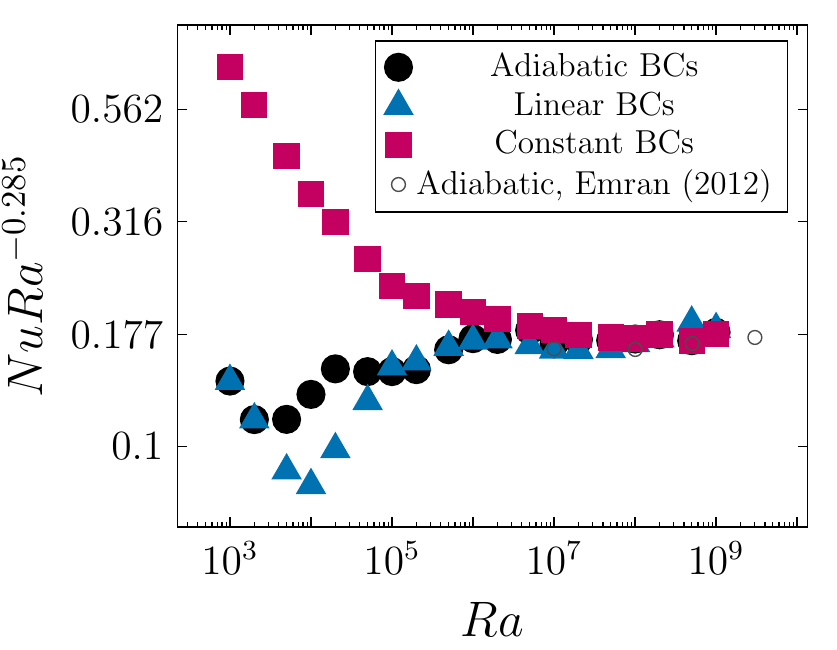}}
\put(0.0, 5.3){$(a)$}
\put(6.8, 5.3){$(b)$}
\put(3.3, 5.4){2D Box}
\put(9.4, 5.4){3D Cylinder}
\end{picture}
\caption{
Nusselt number $\Nu$ for cases with different sidewall boundary conditions in $(a)$ 2D simulations, $(b)$ 3D simulations. For comparison, open symbols shows heat transport in a periodic 2D domain with $\Gamma=2$ by \cite{Johnston2009} $(a)$ and for cylindrical setup with adiabatic sidewalls, $\Gamma=1$ and $Pr=0.7$ conducted by \cite{Emran2012} $(b)$. Dashed lines in $(a)$ show the results from the steady-state analysis.
}
\label{fig:Nu}
\end{figure}

Lastly, the global heat transport is discussed. The results are shown in figure \ref{fig:Nu}. For the 2D setup, we include the results from the steady-state analysis from the first part of this study. Here, we find a very good agreement between $Nu$ of the DNS and steady-states for the SRS mode as well as for the DRS state for adiabatic sidewalls. However, the DRS state for linear sidewalls shows slightly larger $Nu$ in the DNS. This is because the DRS state is an unstable equilibrium solution that can oscillate strongly, which apparently enhances heat transport properties. 

We find that $Nu$ degrades strongly when switching from a SRS- to a DRS-dominated regime at $Ra\approx10^5$ (linear) and $Ra\approx10^6$ (adiabatic) for the 2D domains (figure \ref{fig:Nu}$a$). In contrast, this does not occur for the cylindrical setup as it transitions from the SRS to the double-toroidal state (figure \ref{fig:Nu}$b$). In fact, this flow transition is hardly observed in the evolution of heat transport.

In the high $Ra$ regime, the heat transport in the the cylindrical setup is found to be more efficient than in the 2D setup, with about $30\%$ larger $Nu$. This agrees well with the observations of \cite{Poel2013}. Both setups show $Nu\sim Ra^{0.285}$ scaling at the largest studied $Ra$. We also observe that $Nu$ becomes independent of the choice of sidewall BCs for high $Ra$. This agrees with \cite{Stevens2014}, at least when the sidewall temperature is equal to the arithmetic mean of bottom and top plate temperature. If this condition is violated, \cite{Stevens2014} has shown that $Nu$ differences will exist even for high $Ra$. This indicates that the effects of an imperfectly insulated sidewall tend to be small in experiments when the mean temperature of the sidewall is well controlled.

\subsection{Prandtl number dependence}
\label{sec:pr-dep}

The previous analysis focused on fluids with $Pr=1$, but thermal convection is relevant in nature in a wide variety of fluids and many experiments are conducted in water ($Pr\approx4$) or in liquid metals ($Pr\ll 1$) \citep{Zwirner2020}. Therefore, we now explore the $Pr$ parameter space with $Pr=0.1, 1$ and $10$ for $Ra$ up to $10^9$ in the 2D RBC setup.

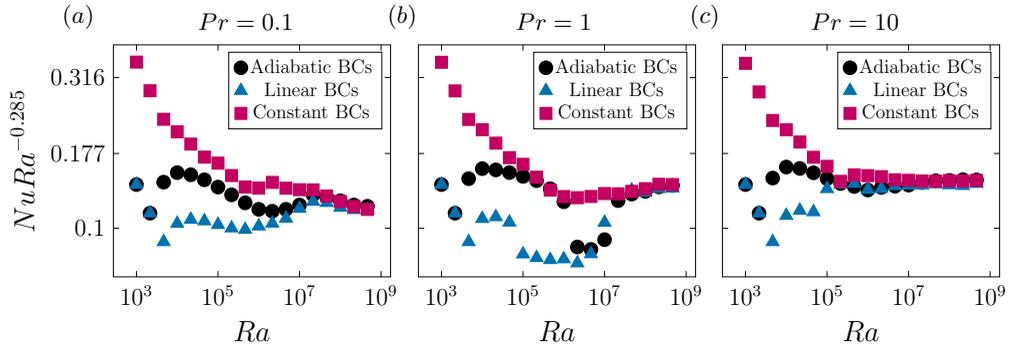
\begin{figure}
\unitlength1truecm
\begin{picture}(12, 4.8)
\put(0.0, 0.0){\includegraphics[height=4.2cm]{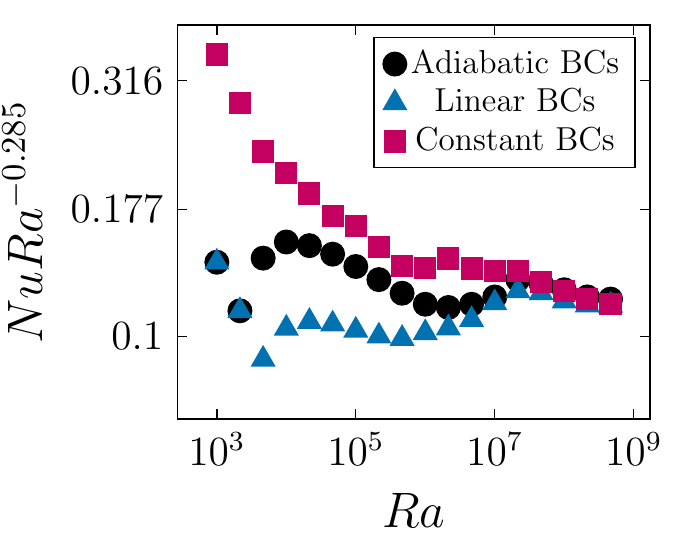}}
\put(5.38, 0.0){\includegraphics[height=4.2cm, trim=50 0 0 0, clip]{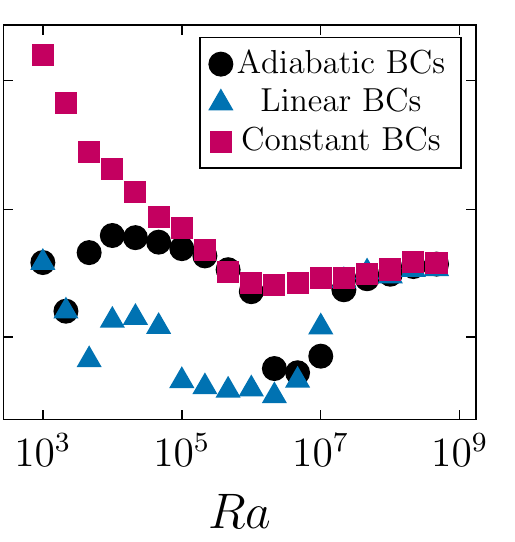}}
\put(9.4, 0.0){\includegraphics[height=4.2cm, trim=50 0 0 0, clip]{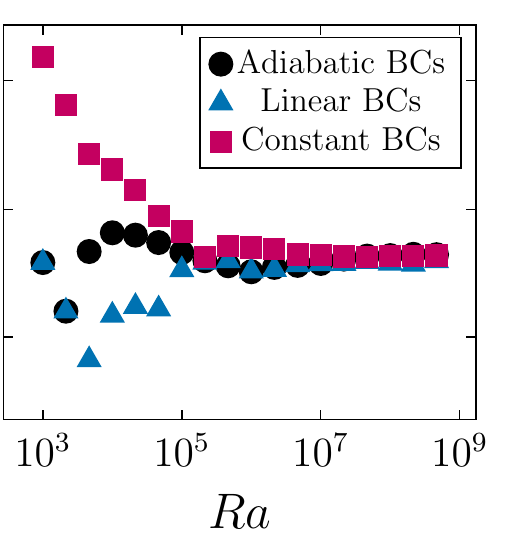}}
\put(0.7, 4.3){$(a)$}
\put(5.0, 4.3){$(b)$}
\put(9.0, 4.3){$(c)$}
\put(2.5, 4.2){$Pr=0.1$}
\put(6.7, 4.2){$Pr=1$}
\put(10.6, 4.2){$Pr=10$}
\end{picture}
\caption{
Nusselt number $Nu$ for $(a)$ $Pr=0.1$, $(b)$ $Pr=1$ and $(c)$ $Pr=10$ in 2D RBC with different thermal sidewall BCs.
}
\label{fig:Nu_Pr}
\end{figure}

The Nusselt number is shown in figure \ref{fig:Nu_Pr}. We observe a collapse of all data points for all studied boundary conditions at large $Ra$. However, the collapse for large $Pr$ is achieved earlier, at $Ra\gtrapprox 10^7$, whereas the differences between $Pr=1.0$ and $Pr=0.1$ are small. Both indicate heat transport invariance for $Ra\gtrapprox 10^8$. This suggests that the size of the thermal boundary layer $\lambda_\theta$ plays a crucial role. For small $Pr$ we expect larger thermal boundary layers, which extend further into the bulk and thus have a stronger influence on the system. As $\lambda_\theta$ gets smaller, the coupling between the sidewall and bulk disappears, and so do the differences in heat transport. And although our results show a small $Pr$-dependence, the main message remains. Experiments with very high $Ra$ are not affected by different thermal sidewall BCs, regardless of whether they are performed in a low $Pr$ or high $Pr$ medium.

\section{Conclusion}
We have investigated the influence of three different lateral thermal boundary conditions, i.e., adiabatic, linearly distributed in the vertical direction and constant (isothermal) ones, on heat transport and flow states in two- and three-dimensional Rayleigh-B\'enard convection (RBC) using direct numerical simulation and steady-state analysis. The steady-state analysis is based on an adjoint-descent method \citep{Farazmand2016}. We found superior convergence chance in the laminar and weakly laminar regime compared to Newton's method, but did not achieve convergence at larger $Ra$. Further studies on the proper boundary conditions, the choice of the energy norm and or a combination with Newton's method are needed to further explore the potential of the method in the study of convective flows.

Investigation of the stability of the single-roll state (SRS) revealed that a linear temperature distribution at the sidewall leads to a premature collapse of the SRS compared to adiabatic BCs. In contrast, the stability of the SRS was enhanced by the introduction of constant temperature sidewall BCs. We find that in 2D and for linear and adiabatic sidewall BCs, the collapse of the SRS is followed by a regime in which the preferred flow state is a double-roll state (DRS), where one roll is located on top of the other. The DRS can be found for adiabatic and linear BCs in the regime $10^6 < Ra \leq 10^7$ and $10^5 \leq Ra \leq 10^7$, respectively, and is associated with suppressed heat transport. The DRS can be stable, it can oscillate periodically with a frequency of $\approx 0.1$ free-fall time unit, or it can be chaotic for larger $Ra$. In 3D cylindrical simulations, a similar flow transition occurs. Imposing linear sidewall BCs leads to the emergence of a double-toroidal structure, that prevails over a wide range of $Ra$, i.e., $10^5 \leq Ra \leq 10^8$. Unlike in 2D, the double-toroidal structure does not lead to a heat transport recession.

We confirmed that the collapse of the SRS in 2D RBC is strongly related to the enlarging of corner rolls. Examining the setup with adiabatic sidewalls, there seem to be two regimes with distinct corner roll growth rates. For small $Ra$, the vorticity balance is dominated purely by diffusion and buoyancy in the form of lateral temperature gradients. In this regime, the size of the corner roll $\delta_{CR}$ grows as $\delta_{CR} \sim Ra^{0.21}$, which is consistent with dimensional analysis. For larger $Ra$, the convective flux starts to be of significance and the growth of the corner roll accelerates to $\delta_{CR} \sim Ra^{0.49}$ before the SRS finally collapses and slowly transforms to the DRS state, undergoing several cycles of flow reversals and restabilization.

Analysis of global heat transport and the flow dynamics have shown that for $Ra\leq 10^8$ there are significant differences between the various sidewall BCs. However, for larger $Ra$ and for various $Pr$ these differences disappear and the different sidewall BCs become globally - in terms of their integral quantities - and dynamically similar. In this context, \cite{Verzicco2008} and \cite{Johnston2009} showed that regardless of imposition of fixed temperature or fixed heat flux at the bottom/top plates, high $Ra$ show similar heat transport. Thus, together with our results, we can conclude that the effects of different boundary conditions, at the sidewalls or at the top/bottom plates, are limited for experiments with high $Ra$. However, there are exceptions. For example, when the sidewall temperature differs from the mean fluid temperature, larger $Nu$ differences can occur \citep{Stevens2014}. Thus, in experiments at high Rayleigh numbers, it appears to be more important to control the mean sidewall temperature than to ensure perfectly insulating conditions. However, close to the onset of convection, the sidewall thermal boundary conditions significantly influence the flow organization and heat transport in the system.

\section*{Acknowledgement}
This work was supported by the Deutsche Forschungsgemeinschaft, grants Sh405/10, Sh405/8, Sh405/7 (SPP 1881 Turbulent ``Superstructures"). The authors also acknowledge Leibniz Supercomputing Centre (LRZ) for providing computing time. 

\section*{Declaration of Interests}
The authors report no conflict of interest.

\begin{appendix}
\section{Heat flux}
\label{app:heat_flux}
The temperature equation for an incompressible fluid in dimensional units is
\begin{align}
\pd{\theta}/\pd t+\nab \cdot {(\bf u\theta)} &= \kappa \nab^2 {\theta}.
\label{eq:T}
\end{align}
Averaging equation $\eqref{eq:T}$ over time yields the following relations for the heat flux $\mathbf{F}$:
\begin{align}
\div\mathbf{F}=0, \quad \mathbf{F}\equiv \uu \theta -\kappa\nab{\theta}.
\label{eq:F_app}
\end{align}
Using the divergence theorem we obtain
\begin{align}
\int_S \mathbf{F} \cdot \mathbf{n} dS=0,
\label{eq:Fs}
\end{align}
which states that the net heat flux through the walls must be zero. Expressing the heat fluxes by the Nusselt number and decomposing the contribution of the surface integral into those for a lower plate heat flux $Nu_h$, for an upper plate heat flux $Nu_c$ and for a side wall heat flux $Nu_{sw}$, we write
\begin{align}
Nu_c - Nu_h + \zeta \langle Nu_{sw} \rangle_z = 0,
\label{eq:Nubalance}
\end{align}
where $\langle \cdot \rangle_z$ denotes a vertical mean and $\zeta$ a geometric factor defining the ratio of the sidewall surface to the bottom/top plate surface, which is $\zeta = 1/\Gamma$ for the 2D box and $\zeta = 4/\Gamma$ for the cylindrical setup. Note that the lateral heat flux $Nu_{sw}$ is $z$-dependent as it was shown in section \ref{sec:dtdr}. For the 2D box this is
\begin{align}
Nu_{sw} = \frac{H}{\Delta} \left[ \frac{\partial \theta}{\partial x}\eval_{x=L} - \frac{\partial \theta}{\partial x}\eval_{x=0} \right]
\label{eq:dtdr_2d_app}
\end{align}
and for the 3D cylinder setup it is
\begin{align}
Nu_{sw} = \frac{H}{2\pi \Delta} \int_0^{2\pi} \frac{\partial \theta}{\partial r}\eval_{r=R} d\varphi.
\label{eq:dtdr_cyl_app}
\end{align}

\section{Thermal dissipation rate}
\begin{figure}
\unitlength1truecm
\begin{picture}(12, 5.)
\put(3.5, 0.2){\includegraphics[width=6.5cm]{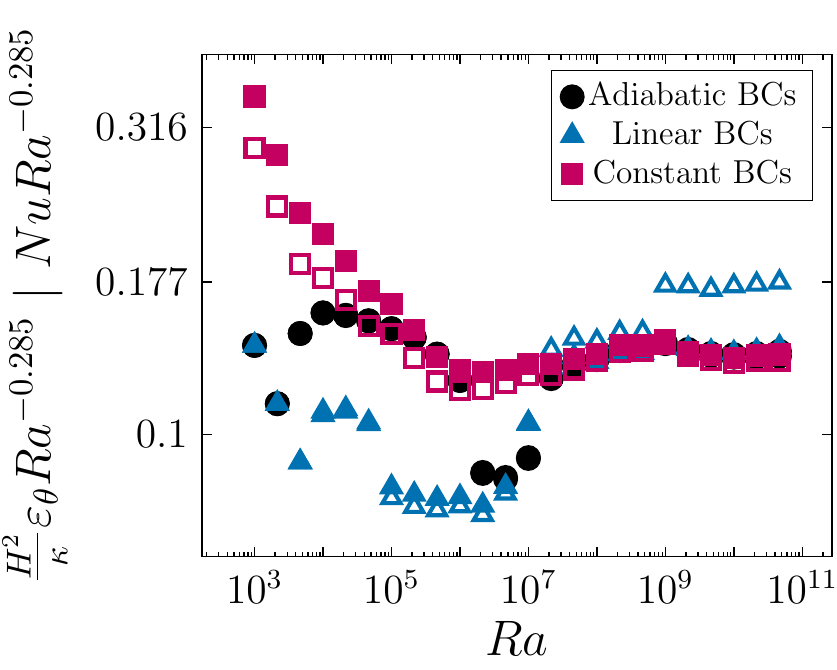}}
\end{picture}
\caption{
Comparison of $Nu$ (closed symbols) and thermal dissipation rate $\varepsilon_\theta$ (open symbols) in the 2D box.  The connection between thermal dissipation and $Nu$ is given in equation \eqref{eq:epsNu}.
}
\label{fig:app1}
\end{figure}

Multiplying equation $\eqref{eq:T}$ with $\theta$ and averaging over time yields
\begin{align}
\frac{1}{2}\pd_t\theta^2 + \frac{1}{2}\nab \cdot {({\bf u}\theta^2)}&= \kappa \theta \nab^2 {\theta}.
\label{eq:T2}
\end{align}
Taking a time and volume average of $\eqref{eq:T2}$, the time derivative and the convective part (for impenetrable walls) vanish and using the relation $(\nab\theta)^2 = \div{(\theta \nab\theta)} - \theta \nab^2 \theta$ we obtain
\begin{align}
\kappa \int_V \overline{(\nab\theta)^2} dV = \kappa \int_V \div{(\overline{\theta \nab\theta})} dV
\label{eq:eth1},
\end{align}
where an overbar denotes a time average and $\varepsilon_\theta = \kappa (\nab\theta)^2$ is known as the thermal dissipation rate. Using the divergence theorem once more, we find the relation between the total thermal dissipation rate and the wall heat fluxes
\begin{align}
\int_V \overline{\varepsilon_\theta} dV = \kappa \int_{S} (\overline{\theta \nab\theta})\cdot \vec{n} dS.
\label{eq:eth}
\end{align}
For clarification, writing eq. $\eqref{eq:eth}$ more explicitly and only for 2D Cartesian coordinates, we get
\begin{align}
\langle\overline{\varepsilon_\theta}\rangle_V &= \frac{\kappa}{V} \left( L \left[ \langle \overline{\theta\partial_z\theta}\rangle_x \right]_{z=0}^{z=H} +H \left[\langle \overline{\theta\partial_x\theta}\rangle_z \right]_{x=0}^{x=L}\right),
\end{align}
with the horizontal and vertical average $\langle\cdot\rangle_x$ and $\langle\cdot\rangle_z$, respectively. In RBC, the temperatures of the upper and lower plates are spatially homogeneous, i.e. $\theta_h=\frac{\Delta}{2}$ and $\theta_c=-\frac{\Delta}{2}$, and assuming that the vertical wall fluxes are equal (which is not necessarily the case for non-adiabatic sidewalls, but has been shown to be true in all our simulations), i.e., $\partial_z\theta_c=\partial_z\theta_h$, then
\begin{align}
\langle\overline{\varepsilon_\theta}\rangle_V &= \frac{\kappa}{V} \left( -L\Delta\langle\partial_z\theta_h\rangle_x+H \left[\langle \overline{\theta\partial_x\theta}\rangle_z \right]_{x=0}^{x=L}\right),\notag\\
\langle\overline{\varepsilon_\theta}\rangle_V &= \frac{\kappa \Delta^2}{H^2} Nu + \frac{\kappa}{L} \left[\langle \overline{\theta\partial_x\theta}\rangle_z \right]_{x=0}^{x=L}.
\label{eq:epsNu}
\end{align}
This results in $\langle\overline{\varepsilon_\theta}\rangle_V = \frac{\kappa \Delta^2}{H^2} Nu$ for adiabatic sidewalls or for zero temperature sidewalls, but adds an additional term to the $\varepsilon_\theta-Nu$ relation otherwise. A comparison of $Nu$ and $\varepsilon_\theta$ is shown in figure \ref{fig:app1}. The virtual discontinuity of $\varepsilon_\theta$ for the linear sidewall temperature reflects the reordering of the flow structures as explained in the main part of this study, but surprisingly $Nu$ shows a rather smooth change in this regime.

\section{Adjoint descent}
\label{app:adjoint}




\subsection{Derivation}
Following \cite{Farazmand2016}, we define the right-hand side of the Navier-Stokes equations as the vector $\mathbf{F_0}$, i.e.
\begin{equation}
    \mathbf{F_0}(\qq) = 
    \begin{pmatrix}
    -\uu \cdot \nab \uu - \nab p + \nu \nab^2 \uu + \vec{\mathbf{e}}_z \theta \\[\jot]
    -\uu \cdot \nab \theta + \kappa \nab^2 \theta \\[\jot]
    \nab\cdot\uu
    \end{pmatrix}.
\label{eq:rhs}
\end{equation}
The functional Gateaux derivative $\delta F(\uu,\uu^\prime) \coloneqq \lim \limits_{\varepsilon \to 0}\frac{F(\uu + \varepsilon \uu^\prime)-F(\uu)}{\varepsilon}$ of equation \eqref{eq:rhs} is
\begin{equation}
    \delta F(\qq,\qq^\prime) = 
    \begin{pmatrix}
    -\uu^\prime \cdot \nab \uu -\uu \cdot \nab \uu^\prime - \nab p^\prime + \nu \nab^2 \uu^\prime + \vec{\mathbf{e}}_z \theta^\prime \\[\jot]
     -\uu^\prime \cdot \nab \theta -\uu \cdot \nab \theta^\prime + \kappa \nab^2 \theta^\prime \\[\jot]
     \nab\cdot\uu^\prime
    \end{pmatrix}.
\label{eq:gateux}
\end{equation}
We want to find the adjoint operator $\delta F^\dagger$ of equation \eqref{eq:gateux} with respect to the inner-product
\begin{equation}
\langle\qq,\qq^\prime \rangle_\mathcal{A} = \int_\mathcal{D} \left( \qq \cdot \mathcal{A} \qq^\prime \right) \text{d}\bf x .
\label{eq:inner}
\end{equation}
The adjoint $\delta F$ of equation \eqref{eq:gateux} with respect to the inner product \eqref{eq:inner}, with $\tilde{\qq}\equiv \mathcal{A}\qq$, is derived as follows
\begin{align}
\langle \delta F(\qq,\qq^\prime), \tilde{\qq}\dprime \rangle_{\mathcal{A}} &= \nonumber\\
&= \int_V
\begin{pmatrix}
    -\uu^\prime \cdot \nab \uu -\uu \cdot \nab \uu^\prime - \nab p^\prime + \nu \nab^2 \uu^\prime + \vec{\mathbf{e}}_z \theta^\prime \\[\jot]
     -\uu^\prime \cdot \nab \theta -\uu \cdot \nab \theta^\prime + \kappa \nab^2 \theta^\prime \\[\jot]
     \nab\cdot\uu^\prime
\end{pmatrix}
\begin{pmatrix}
\tilde{\uu}\dprime \\[\jot]
\tilde{\theta}\dprime \\[\jot]
\tilde{p}\dprime
\end{pmatrix}
\text{d}\bf x \nonumber\\[10pt]
&= \int_V
\begin{pmatrix}
    \left(\nab \tilde{\uu}\dprime +\nab \tilde{\uu}^{\prime\prime\text{T} } \right) \uu + \theta\nab \tilde{\theta}\dprime
    - \nab \tilde{p}\dprime + \nu \nab^2 \tilde{\uu}\dprime \\[\jot]
     \uu \cdot \nab \tilde{\theta}\dprime + \nu \nab^2 \tilde{\theta}\dprime +  \vec{\mathbf{e}}_z\cdot \tilde{\uu}\dprime \\[\jot]
     \nab\cdot\tilde{\uu}\dprime
\end{pmatrix}
\begin{pmatrix}
\uu\sprime \\[\jot]
\theta\sprime \\[\jot]
p\sprime
\end{pmatrix}
\text{d}\bf x \nonumber\\[10pt]
&=\langle \qq\sprime, \delta F^\dagger(\qq,\tilde{\qq}\dprime) \rangle_{\mathcal{A}},
\label{eq:adjoint}
\end{align}
where the second line follows from integration by parts. Here we have refrained from writing the boundary terms that follow from the integration by parts step, since they can be eliminated by choosing the boundary conditions on $\tilde{\qq}\dprime$ as discussed in section \ref{sec:adj_descent}.

\subsection{Choice of the norm}
\label{app:norm}
As mentioned in \cite{Farazmand2016}, the most obvious choice for the norm is the $\text{L}^2$ norm, i.e. $\mathcal{A}=I$, where $I$ is the identity operator. However, this norm is rather stiff and leads to restrictive small time steps. As an alternative, \cite{Farazmand2016} uses a norm related to the Laplacian, which effectively smooths the $\tilde{\qq}\dprime$ field. Here we use a similar norm based on the inversed Laplacian, i.e. $\mathcal{A} = (I - \alpha \nab^2)^{-1}$,
\begin{equation}
\langle\qq,\qq^\prime \rangle_{\nab^{-2}} = \int_V \left( \qq \cdot \mathcal{A} \qq^\prime \right) \text{d}\bf x = \int_V \left( \qq \cdot \tilde{\qq}^\prime  \right) \text{d}\bf x
\label{eq:norm}
\end{equation}
where $a$ is a positive constant. Then, $\tilde{\qq}^\prime$ is obtained as the solution of the Helmholtz equation
\begin{equation}
(I - \alpha \nab^2)\tilde{\qq}^\prime = \qq^\prime,
\end{equation}
which points out the smoothing property of this norm. In practice, we choose $\alpha=1$. The choice of the operator for the energy norm is somewhat arbitrary, but this peculiar choice leads to improved numerical stability properties. Note that the operator $\mathcal{A}$ should be positive definite and should commute with the divergence operator, i.e. $\mathcal{A}(\nab\cdot\uu) = \nab\cdot\mathcal{A}\uu$.

\FloatBarrier

\end{appendix}

\bibliographystyle{jfm}
\bibliography{References}
\end{document}